\RequirePackage{lineno}

\documentclass[aps,prc,twocolumn,linenumbers,superscriptaddress,showpacs,floatfix]{revtex4}
\usepackage{graphicx}
\usepackage{dcolumn}
\usepackage{bm}
\usepackage{xcolor}
\usepackage{lineno}

\usepackage{subfigure}
\usepackage{amsmath,epsfig}
\usepackage{epstopdf}
\usepackage{multirow}
\usepackage{hhline}
\usepackage{afterpage}
\usepackage{float}
\usepackage{hyperref} 
\usepackage{booktabs}

\newcommand{\pt}{p_T}
\newcommand{\Ihadrons}{\pi^{\pm}, K^{\pm}, p, \mathrm{and}~\bar{p}}
\newcommand{\Elabs}{E_{lab} = 6.7,~8,~11,~\mathrm{and~25~A~GeV}}
\newcommand{\snn}{\sqrt{s_{NN}}}
\newcommand{\ratios}{\pi^{-}/\pi^{+}, K^{-}/K^{+},~\mathrm{and}~\bar{p}/p}

\newcommand{\meanpt}{\langle p_T \rangle}
\newcommand{\npart}{N_{part}}

\newcommand{\midr}{|y| < 0.5}

\begin{document}
\title{Beam energy dependence of identified particle production in heavy-ion collisions using a parton-hadron string dynamics model}

\author{T. Bhat}
\affiliation{Department of Physics, Panjab University, Chandigarh 160014, India}

\author{V. Bairathi}
\affiliation{Instituto de Alta Investigación, Universidad de Tarapacá, Casilla 7D, Arica 1000000, Chile}

\author{L. Kumar}
\affiliation{Department of Physics, Panjab University, Chandigarh 160014, India}

\author{S. Kabana}
\affiliation{Instituto de Alta Investigación, Universidad de Tarapacá, Casilla 7D, Arica 1000000, Chile}

\date{\today}

\begin{abstract}
We report predictions for the transverse momentum ($p_T$) spectra of $\pi^{\pm}$, $K^{\pm}$, $p$, and $\bar{p}$ in various collision centralities from Au + Au collisions at beam energies ($E_{lab}$) of 6.7, 8, 11, and 25 A~GeV using a parton-hadron string dynamics (PHSD) transport model. We studied the dependence of particle yields ($dN/dy$), mean transverse momenta ($\langle p_T \rangle$), and particle ratios on collision energy and centrality to understand the underlying mechanisms of particle production. A comparison of the PHSD model results with available experimental measurements provides a qualitative description of these observables. Our results highlight the importance of baryon stopping, strangeness production, pair production, and baryon-antibaryon annihilation in the high baryon density region. These findings also provide theoretical insights relevant to the ongoing beam energy scan program at RHIC and the future heavy-ion programs at FAIR and NICA.
\end{abstract}
\pacs{25.75.Dw}
\maketitle

\section{Introduction}
\label{sec:Intro}
A key objective of heavy-ion collision experiments is to explore the Quantum Chromodynamics (QCD) phase diagram. This diagram is generally represented with temperature ($T$) and baryon chemical potential ($\mu_{B}$). In the context of heavy-ion collisions, if we assume that a thermalized system is formed, the parameters $T$ and $\mu_{B}$ can be varied by changing the collision energy~\cite{r1,r2,r3}. Theoretical calculations from lattice QCD predict a phase transition from hadronic to a deconfined matter of  quarks and gluons known as the Quark-Gluon Plasma (QGP) at high temperature and high baryonic densities~\cite{r4,r5,r6}.

Various experimental observations suggest the formation of a QGP medium and a possible phase transition. These observations can be broadly classified into two types: rare and bulk probes. The rare probes include direct photon and dilepton production, as well as jet modification. These are considered more robust indicators because of their minimal interactions with the final state. However, their detection is challenging due to low production rates. On the other hand, bulk probes, such as the transverse momentum ($\pt$) spectra of produced particles, enhanced strangeness production, increased antibaryon yields, and strong collective flow, are more accessible but may be significantly influenced by final-state interactions~\cite{r7}.

 A series of experimental studies has explored various aspects of hadron production and transport dynamics across different collision energies and systems. The beam energy dependence of identified hadrons ($\Ihadrons$) yield in Au + Au collisions at $\snn = 7.7 - 39$ GeV from the STAR beam energy scan (BES) program suggests a change in particle production mechanism below 19.6 GeV~\cite{r8}. The results indicate more baryon stopping at lower beam energies, and the behavior of the $K/\pi$ ratio suggests that the maximum baryon density in these collisions is attained around $\snn = 7.7$ GeV~\cite{r8}. In an another study in Au + Au collisions at $\snn = 7.7 - 39$ GeV, the results highlight significant energy-dependent features in strange hadron production and suggest that the region below $\snn = 19.6$ GeV is critical for exploring the onset of de-confinement~\cite{r9}. A comparative study of identified hadrons from a transport model, using different initial conditions and particle production mechanisms, has highlighted the need to consistently describe the bulk properties of the system formed in heavy-ion collisions across various energies and centralities~\cite{r9b}.
 
Measurements of bulk probes for identified hadrons in Au + Au collisions at $\snn = 3$ GeV from the STAR Fixed-Target (FXT) program show that the dominant particle production mechanism is different than that at high energies. Furthermore, at higher energies, the Grand Canonical Ensemble (GCE) description is necessary to explain the experimental data, while at lower energies, the Canonical Ensemble (CE) is required~\cite{r10,r11,r12}. Measurements of pion production in Au + Au collisions at $\snn =$ 2.4 GeV from the HADES experiment show that pion production increases moderately with decreasing centrality and system size~\cite{r13}.  

Heavy-ion collisions at $\snn\sim$ 2$-$10~GeV produce matter at baryon densities comparable to those found in neutron star interiors~\cite{r13b}, a regime optimally accessible at FAIR and NICA, where the CBM and MPD experiments are designed to perform high-precision measurements~\cite{r14,r15}. In particular, the CBM experiment will explore beam energies $E_{lab}\sim$ 2$-$11~A~GeV ($\mu_{B}\simeq$~540$-$800~MeV), where net-baryon densities up to 5$-$8 times the normal nuclear density are expected. This enables enhanced sensitivity to in-medium effects, the nuclear equation of state, and possible signatures of de-confinement and chiral symmetry restoration~\cite{r14}. In this context, studying the beam energy dependence of hadron yields will provide benchmarks for future CBM and MPD measurements and serve as a critical tool for mapping the high $\mu_{B}$ region of the QCD phase diagram.

These studies demonstrate progress in understanding the particle production mechanism, the role of in-medium effects, and the sensitivity of observables to model parameters in the low beam energy and high baryon density regime. However, challenges remain in consistently describing all aspects of particle production across different energies and centralities, necessitating further theoretical and experimental efforts.

In this paper, we report the study of $\pt$-spectra for identified hadrons ($\Ihadrons$) at mid-rapidity ($\midr$) in Au + Au collisions at $\Elabs$ ($\snn \approx 4 - 7$ GeV) using the PHSD model. The energy range studied in this work overlaps with the collision energy range of the upcoming experiments at FAIR and NICA. In this beam energy range, the baryon chemical potential is high, and thus, maximal baryon stopping is expected. Additionally, the transverse momentum distributions, hadron yields, and particle ratios are anticipated to exhibit strong sensitivity to the equation of state. There is also a potential onset of de-confinement within this energy range~\cite{r14,r15}. The PHSD model is particularly well-suited for a systematic study of bulk observables in this energy domain, as it offers a unified transport description of both hadronic and partonic dynamics. This enables a consistent treatment of baryon stopping, strangeness production, and final state interactions that influence the measured bulk observables.

The paper is structured as follows: Section \ref{sec:model} introduces the PHSD model and outlines the methodology used in the analysis. Section \ref{sec:results} presents the results on transverse momentum ($p_T$) spectra, integrated yield, average transverse momentum, and particle ratios of identified hadrons in Au + Au collisions at $\Elabs$. Finally, Section \ref{sec:conclusion} summarizes and discusses the findings presented in this paper.

\section{PHSD model}
\label{sec:model}
The PHSD model is a microscopic covariant dynamical transport approach developed to study strongly interacting matter in relativistic heavy-ion collisions~\cite{r16,r17,r18,r19}. It is based on the Dynamical Quasi-Particle Model (DQPM) and formulated within the Kadanoff–Baym framework for Green’s functions in phase space~\cite{r20,r21,r22,r23}. The PHSD model comprehensively describes the evolution of relativistic heavy-ion collisions, from initial hard scatterings and string formation to the dynamical de-confinement transition into a strongly interacting QGP, followed by hadronization and interactions in the expanding hadronic phase. The model incorporates both partonic and hadronic degrees of freedom via off-shell transport equations, with scalar mean fields playing a crucial role in generating collective flow in the partonic phase. A realistic equation of state constrained by lattice QCD (lQCD) calculations is employed~\cite{r24,r25}, and hadronization proceeds through the fusion of off-shell partons into off-shell hadronic states while strictly conserving energy, momentum, and quantum numbers.

In PHSD, nucleus–nucleus collisions begin with hard scatterings among incoming nucleons, modeled using a nonlinear $\sigma$–$\omega$ nuclear equation of state with the NL1 parametrization~\cite{r26}. Color-neutral strings, generated according to the FRITIOF Lund model and implemented via PYTHIA 6.4~\cite{r27,r28,r29}, fragment into effective quarks and antiquarks when the local energy density exceeds the critical value for de-confinement ($\sim$0.5 GeV/$fm^{3}$, as indicated by lQCD). The transition from partonic to hadronic matter is governed by covariant transition rates that convert quark–antiquark pairs into mesons and three-quark clusters into baryons, ensuring detailed balance and entropy production due to the off-shell nature of the degrees of freedom. At lower energy densities, PHSD smoothly reduces to the Hadron–String Dynamics (HSD) model~\cite{r30}. The model also incorporates key aspects of chiral symmetry restoration through the Schwinger mechanism~\cite{r31}, accounting for the in-medium behavior of the scalar quark condensate at high temperature and baryon density. The PHSD model has been extensively validated against experimental data from SPS to RHIC energies~\cite{r32,r33,r34,r35,r36,r37,r38,r39,r40,r41}.

\begin{figure*}[!htbp]
\begin{center}
\includegraphics[scale=0.4]{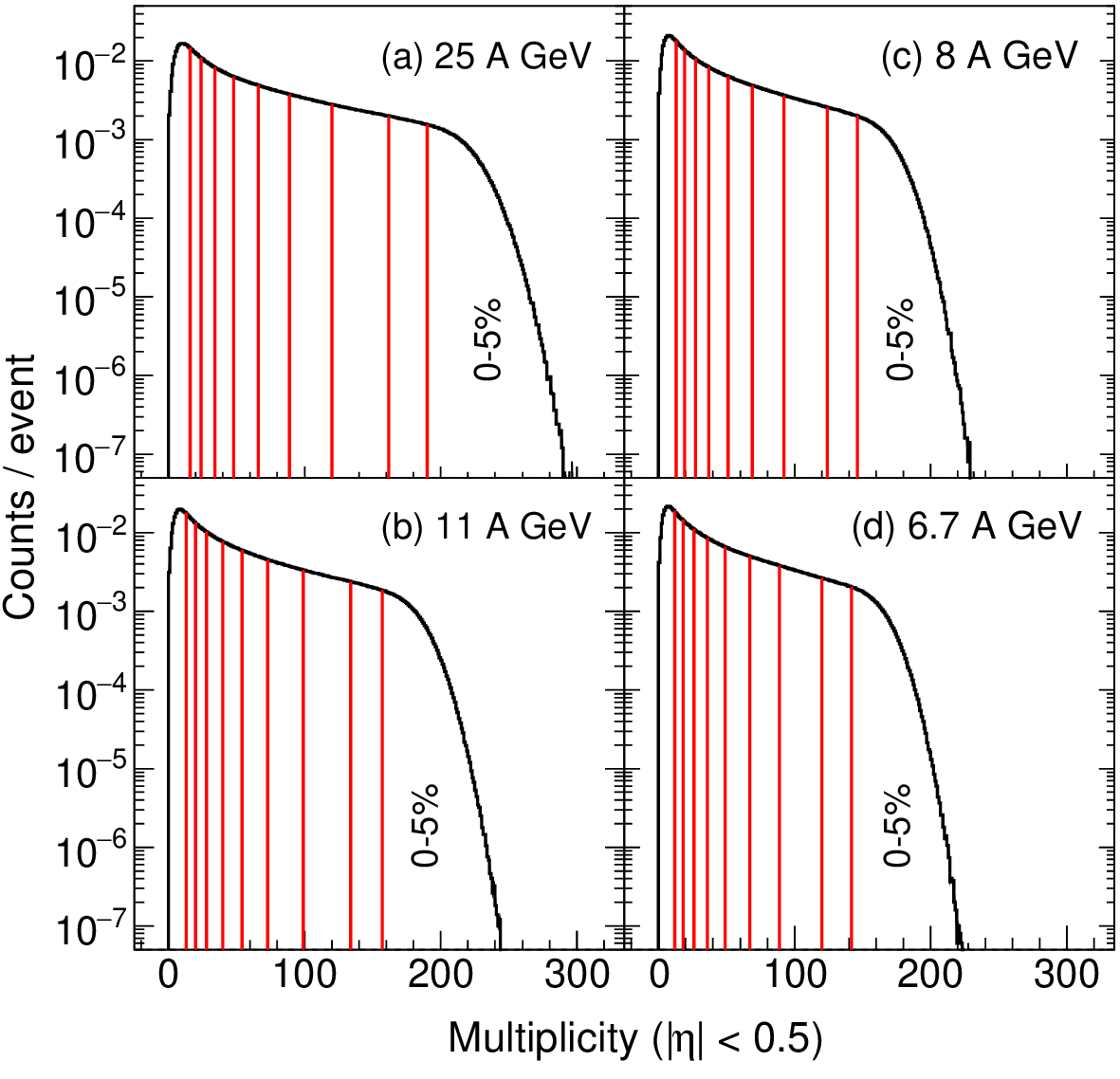}
\caption{(Color online) Multiplicity distribution within $|\eta| <$ 0.5 in Au + Au collisions at $\Elabs$ from the PHSD model. Different centrality classes from 0-5\% to 70-80\% are shown as distinct bands.}
\label{fig:cent}
\end{center} 
\end{figure*}

The results of this study are based on version 4.1 of the PHSD model. A dataset consisting of 50 million minimum-bias Au + Au collision events was generated at beam energies $\Elabs$. The $p_T$-spectra of identified hadrons ($\Ihadrons$) at mid-rapidity ($\midr$) across different centrality classes are obtained using the PHSD model. The centrality classes are determined based on the fractions of the total charged particle multiplicity. The multiplicity refers to the total number of charged particles within a pseudo-rapidity window of $|\eta| < 0.5$. We considered nine centrality classes: 0$-$5\%, 5$-$10\%,  10$-$20\%, 20$-$30\%, 30$-$40\%, 40$-$50\%, 50$-$60\%, 60$-$70\%, and 70$-$80\%, from central to peripheral collisions, as presented in Fig.~\ref{fig:cent}.

\section{Results}
\label{sec:results}
In this section, we present results on transverse momentum spectra of identified hadrons ($\Ihadrons$) at mid-rapidity ($\midr$) in Au + Au collisions at $\Elabs$ using the PHSD model. We report the particle yields and average transverse momentum extracted from the $p_T$-spectra. We also report various particle ratios as a function of the number of participating nucleons ($\npart$). We discuss the beam energy dependence of particle and antiparticle yields and compare them with available experimental data.

\subsection{Transverse momentum spectra}
\label{sec:Pt_spectra}
\begin{figure*}
\begin{center}
\includegraphics[scale=0.6]{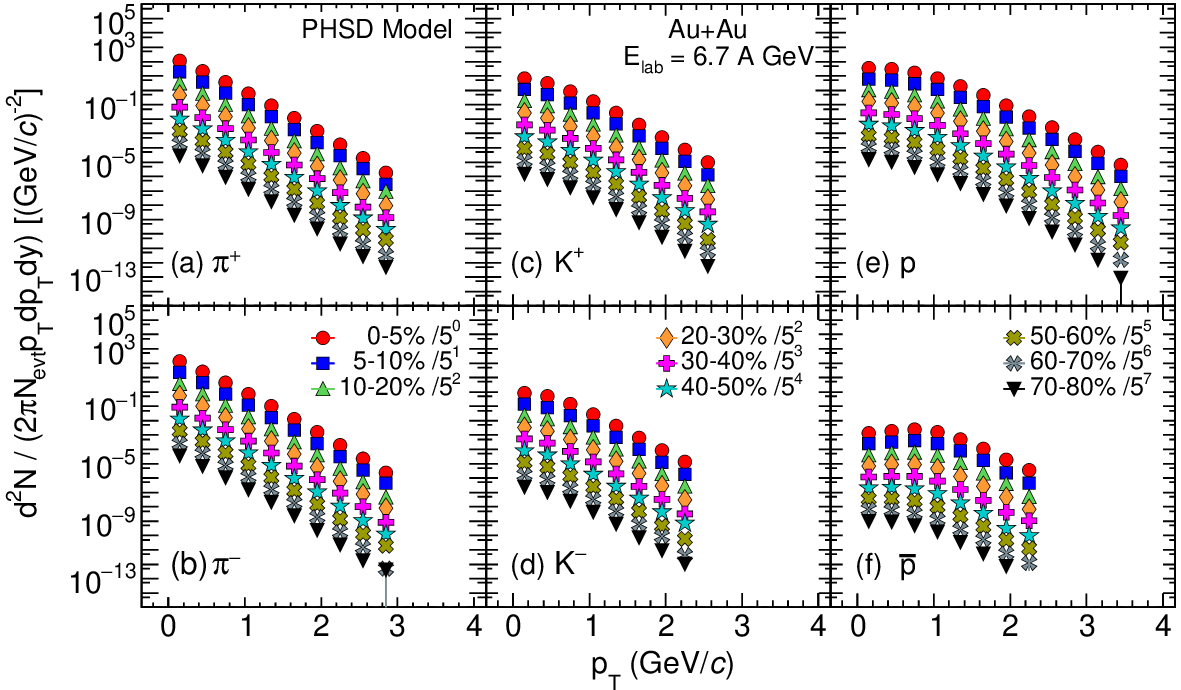}
\caption{(Color online) Transverse momentum spectra of $\Ihadrons$ at $\midr$ in Au + Au collisions at $E_{lab} =$ 6.7 A~GeV for centrality classes 0$-$5\% to 70$-$80\% from the PHSD model. The $\pt$-spectra for different centrality classes are scaled for clarity, as shown in the legends.}
\label{fig:PtSpectra6.7}
\end{center} 
\end{figure*}

\begin{figure*}
\begin{center}
\includegraphics[scale=0.6]{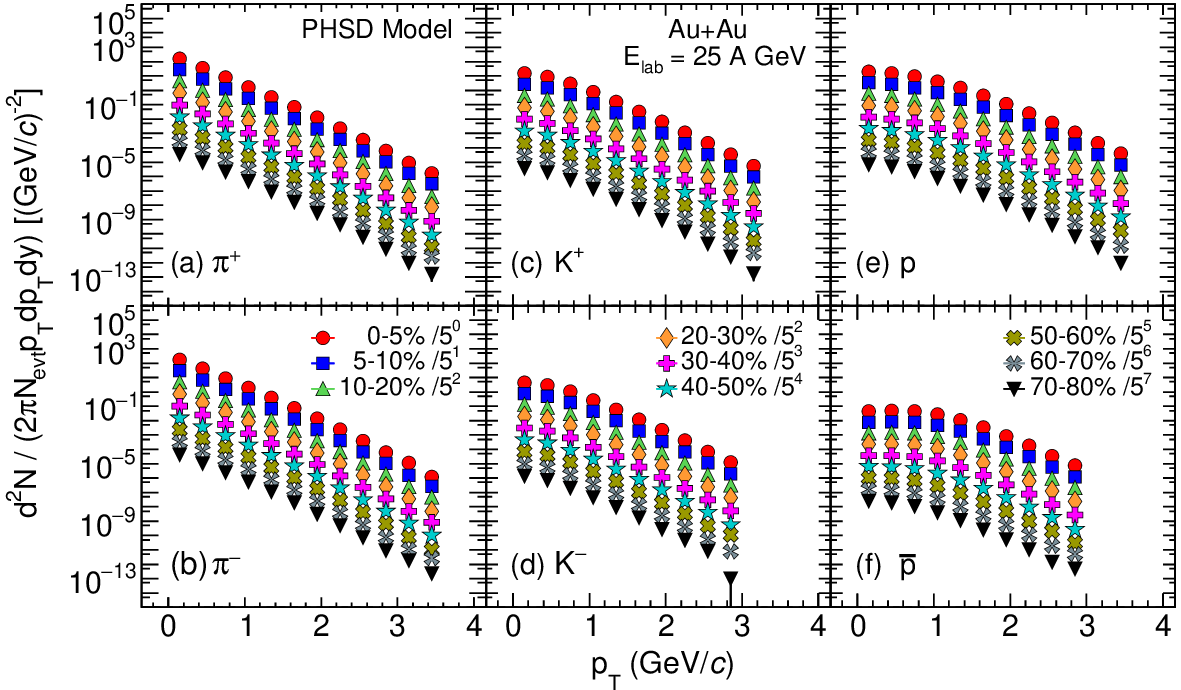}
\caption{(Color online) Same as Fig.~\ref{fig:PtSpectra6.7} but for Au + Au collisions at $E_{lab} =$ 25 A~GeV from the PHSD model.}
\label{fig:PtSpectra25}
\end{center} 
\end{figure*}

The $\pt$-spectra of $\Ihadrons$ in Au + Au collisions at beam energies $E_{lab} =$ 6.7 and 25 A GeV for different collision centralities from the PHSD model are shown in Fig.~\ref{fig:PtSpectra6.7} and Fig.~\ref{fig:PtSpectra25}. For clarity, the spectra have been scaled by successive powers of five for each centrality class. The $\pt$-spectra for other beam energies ($E_{lab} $ = 8 and 11 A GeV) are also obtained but not shown here. For all particle species, the $\pt$-spectra decrease with increasing $\pt$ and from central to peripheral collisions. Additionally, the $\pt$-spectra show a clear dependence on beam energy. The invariant yields of particles, except protons, decrease with decreasing beam energy across all centrality classes. Conversely, the invariant yield of protons increases with decreasing beam energy. A more quantitative characterization of the observed trends is obtained by studying the integrated particle yields ($dN/dy$) and the mean transverse momentum ($\meanpt$), which will be discussed in the following sub-sections.

\subsection{Particle yields}
\label{subsec3}
\begin{figure*}[!htbp]
\begin{center}
\includegraphics[scale=0.35]{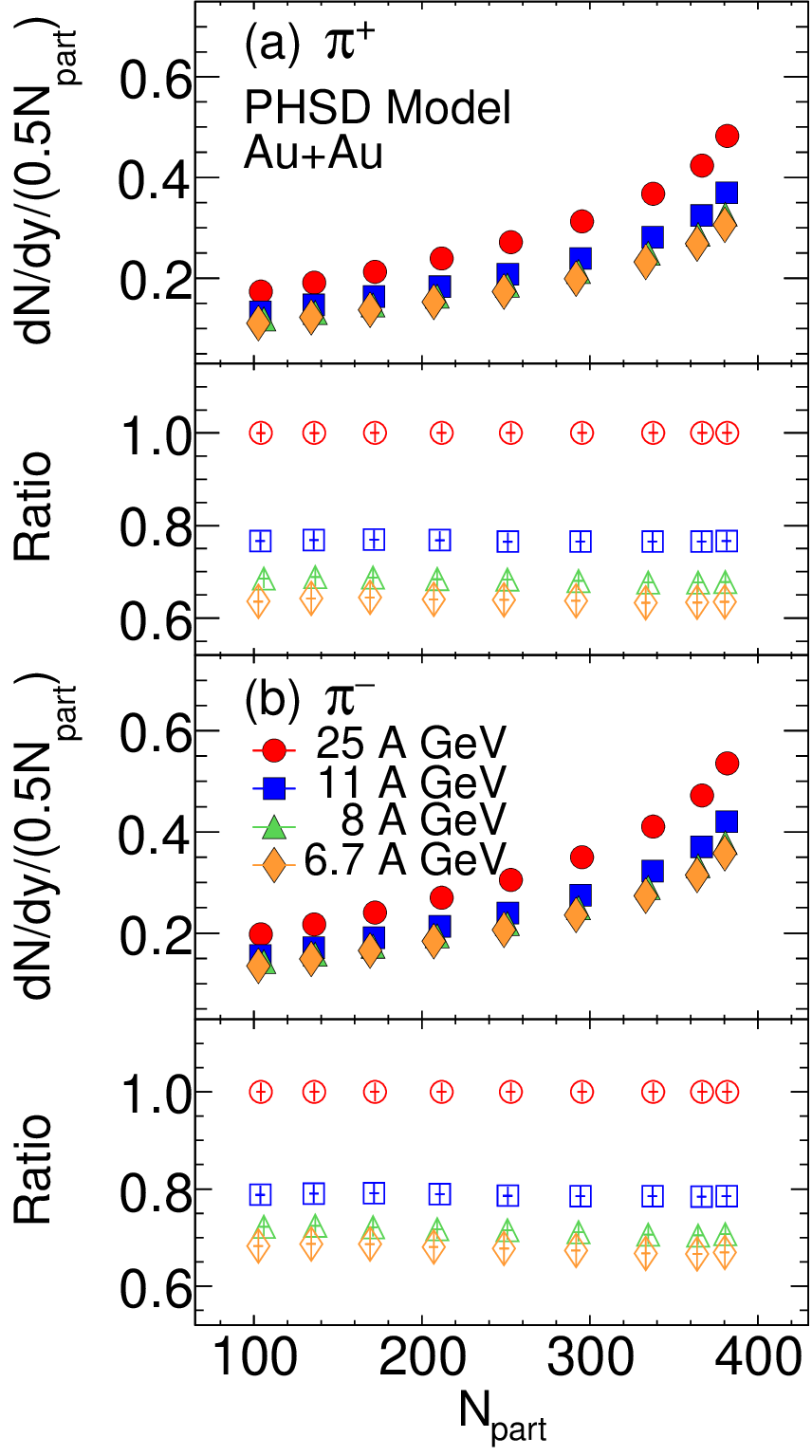}
\includegraphics[scale=0.35]{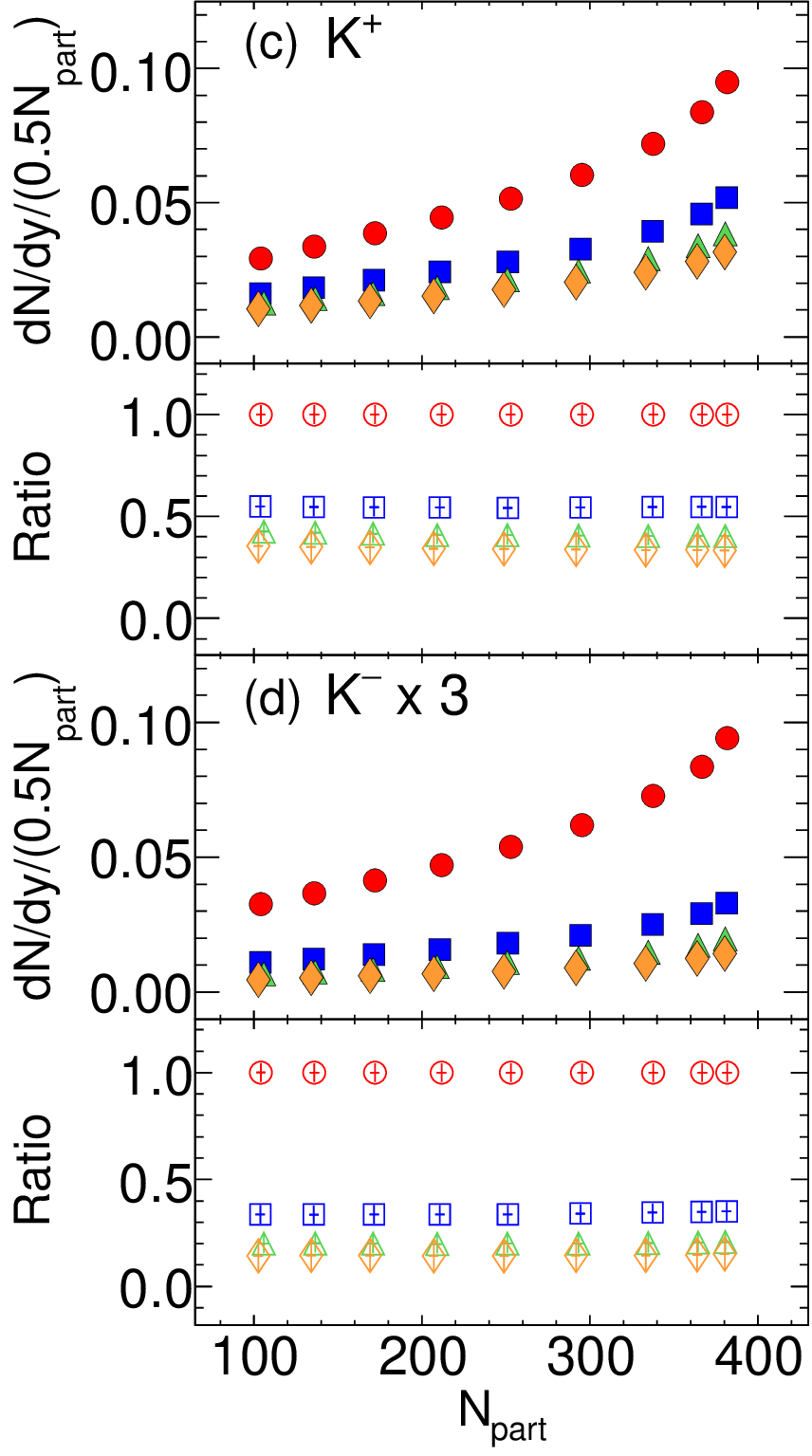}
\includegraphics[scale=0.35]{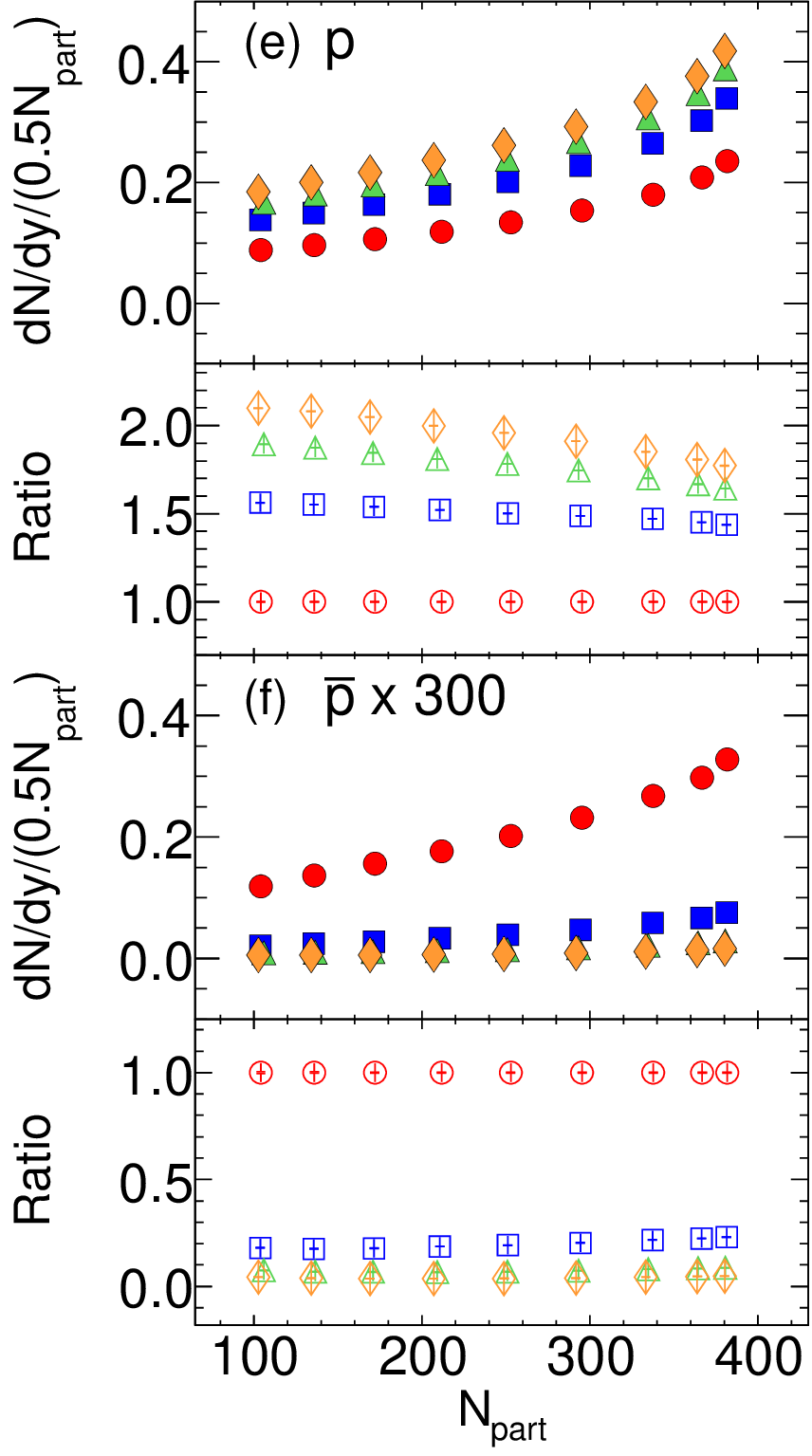}
\caption{(Color online) Particle yields ($dN/dy$) scaled by $\npart$/2 as a function of $\npart$ for $\Ihadrons$ at $\midr$ in Au + Au collisions at $\Elabs$ from the PHSD model. The yields of $K^{-}$ and $\bar{p}$ are scaled by constant factors for each energy to enhance visualization. The bottom panels show the ratios of $dN/dy$ at each beam energy relative to the 25~A~GeV.}
\label{fig:dndyNpart}
\end{center} 
\end{figure*}

The invariant yields ($dN/dy$) of $\Ihadrons$ at $\midr$ as a function of $\npart$ in Au + Au collisions at beam energies $\Elabs$ from the PHSD model are presented in Fig.~\ref{fig:dndyNpart}. The $dN/dy$ values are obtained by counting the number of particles within the rapidity interval of $|y| <$ 0.5. A clear centrality dependence of $dN/dy$ is observed, with the yields decreasing from central to peripheral collisions at a given beam energy. However, the yields of $K^{-}$ and $\bar{p}$ exhibit a weaker dependence on centrality for beam energies below $E_{lab} =$ 25 A~GeV.

The $dN/dy$ values for pions, kaons, and anti-protons decrease with decreasing beam energy. However, the $dN/dy$ values for protons show an increasing trend with decreasing beam energy. The ratios of $dN/dy$ at a given beam energy to the 25 A~GeV for all the particles are shown in the bottom panels of Fig.~\ref{fig:dndyNpart}. For pions, these ratios exhibit a relatively weaker energy dependence compared to those of kaons and protons. This suggests that the production of pions is less sensitive to variations in beam energy within this energy range.

The $dN/dy$ of kaons decreases significantly with decreasing energy, with $K^{-}$ exhibiting a stronger suppression than $K^{+}$, consistent with the dominance of associated production in strangeness at lower beam energies and the reduced contribution from pair production~\cite{r8}. The $dN/dy$ of protons increases with decreasing energy, opposite to the $\bar{p}$. This behavior indicates the accumulation of initial-state protons at mid-rapidity due to enhanced baryon stopping as the beam energy decreases within the PHSD model framework. The significant decrease in $K^{-}$ and $\bar{p}$ yields with beam energy indicates the difference between particles composed of only produced quarks and those containing both produced and transported quarks from colliding nucleons ($\pi^{\pm}$, $K^{+}$, and $p$). Understanding this difference is essential to explore particle production mechanisms in heavy-ion collisions at high baryon density.

\subsection{Mean transverse momentum}
\begin{figure*}[!htbp]
\begin{center}
\includegraphics[scale=0.35]{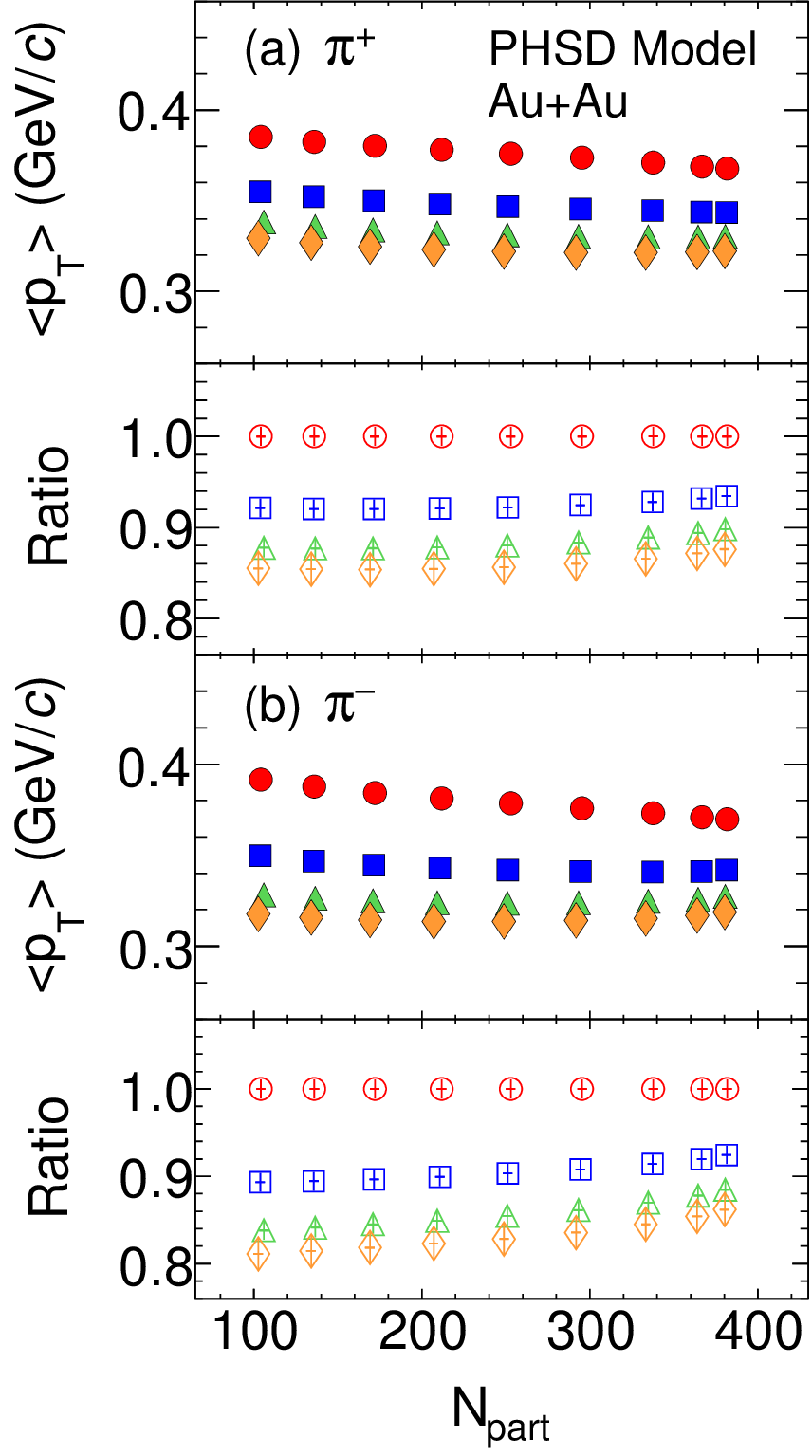}
\includegraphics[scale=0.35]{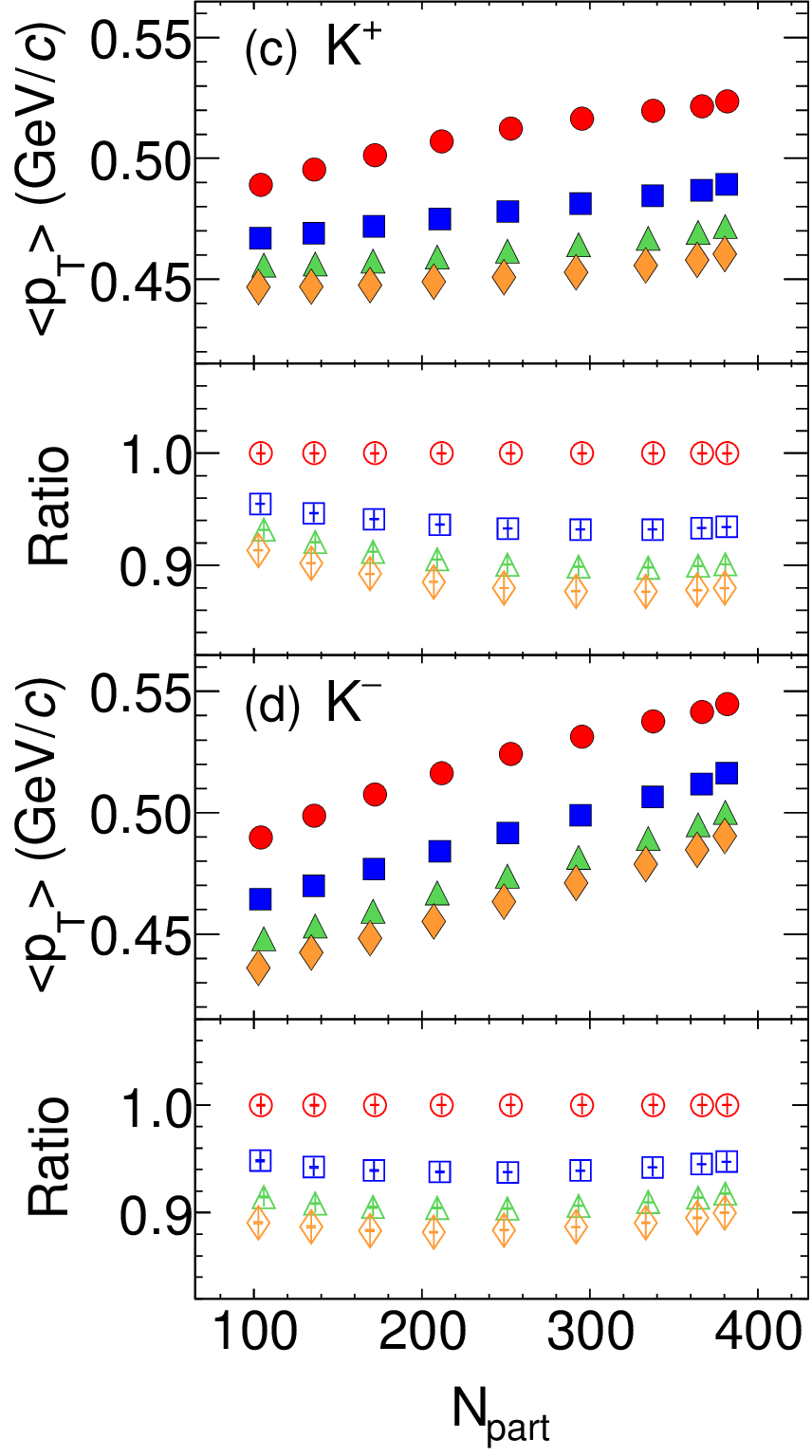}
\includegraphics[scale=0.35]{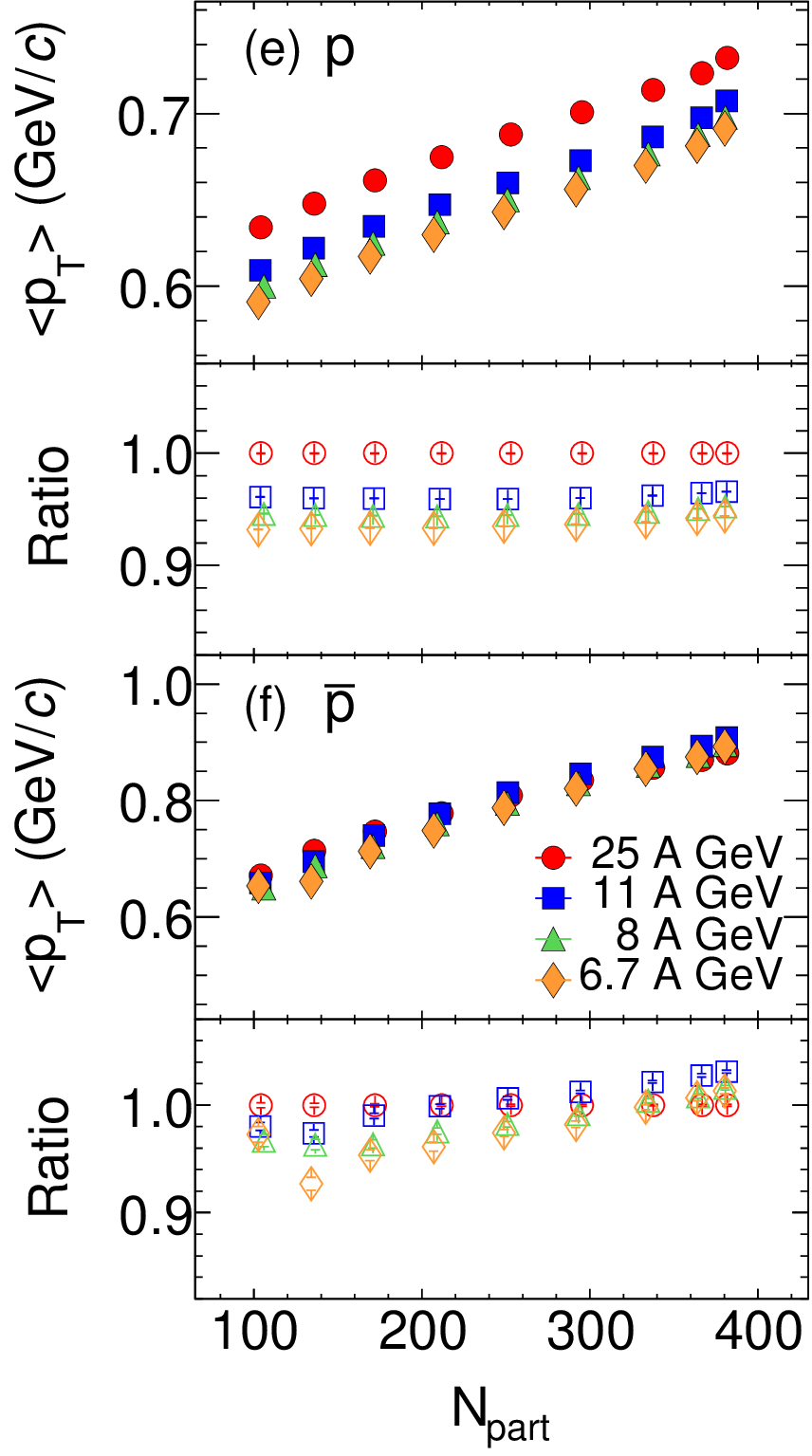}
\caption{(Color online) Mean transverse momentum ($\meanpt$) for $\Ihadrons$ as a function of $\npart$ at $\midr$ in Au + Au collisions at $\Elabs$ from the PHSD model. The bottom panels show the ratios of $\meanpt$ at each beam energy relative to the 25 A GeV.}
\label{fig:meanPt}
\end{center} 
\end{figure*}

Figure~\ref{fig:meanPt} shows mean transverse momentum ($\meanpt$) for $\Ihadrons$ as a function of $\npart$ at $\midr$ in Au + Au collisions at beam energies $\Elabs$ from the PHSD model. For all the particles, the $\meanpt$ is obtained from their $p_{T}$-distribution. The values of $\meanpt$ increase with the particle mass at all beam energies. For pions, $\meanpt$ exhibits a weak dependence on $\npart$ across all energies, suggesting that the transverse dynamics of pions are only slightly affected by collision centrality within the energy range of $E_{lab} =$ 6.7 to 25 A GeV. In contrast, kaons and protons show an apparent decrease in $\meanpt$ from central to peripheral collisions. This trend reflects a decrease in collective transverse expansion as collisions become more peripheral, with heavier particles being affected more due to their larger masses. The ratios of $\meanpt$ for each particle at different beam energies with respect to 25 A~GeV are presented in the bottom panels of Fig.~\ref{fig:meanPt}. The $\meanpt$ decreases with a decrease in beam energy for all particles, which further indicates the reduction in collective dynamics at lower beam energies.  

\subsection{Particle ratios}
\begin{figure*}[!htbp]
\begin{center}
\includegraphics[scale=0.29]{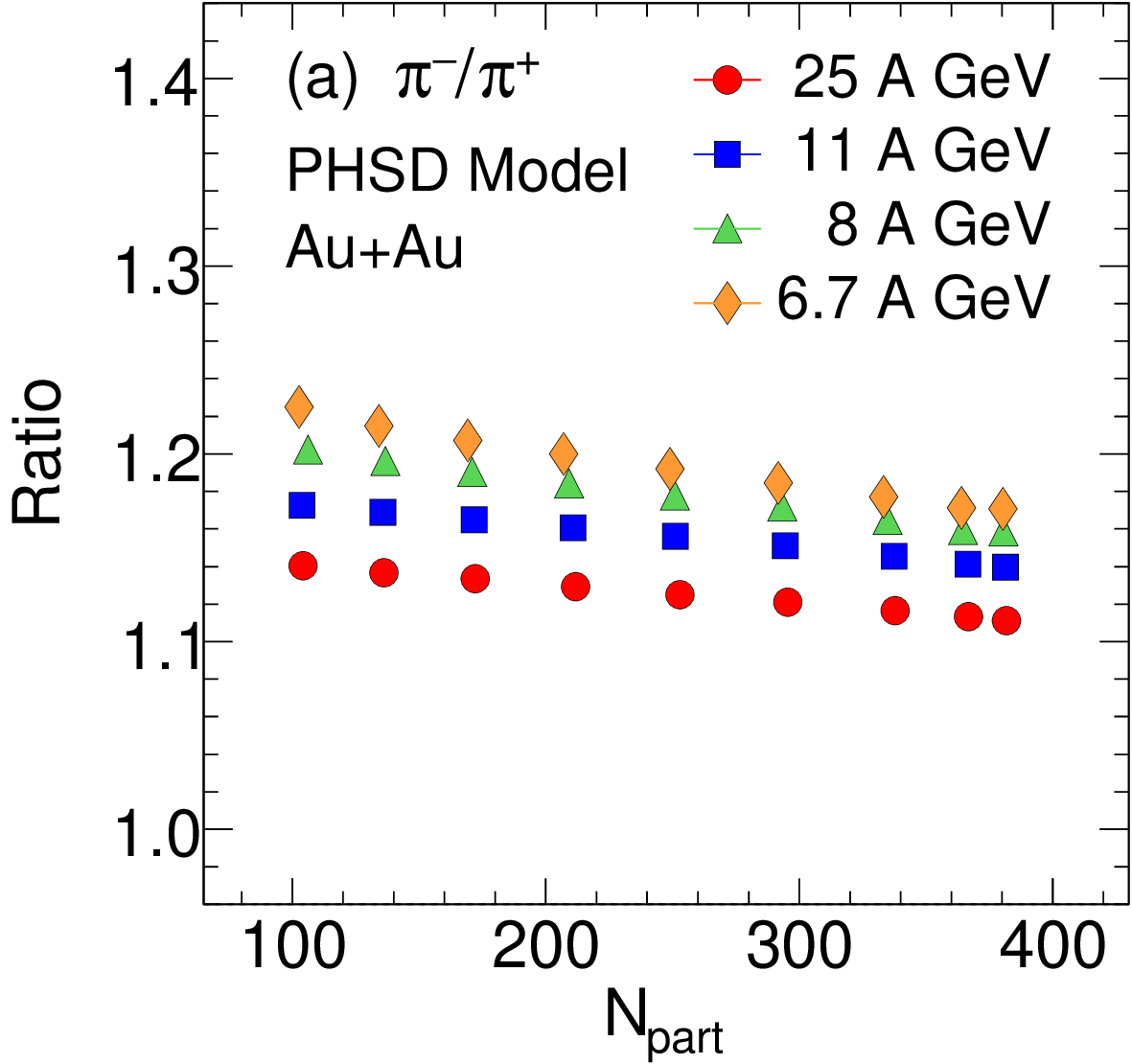}
\includegraphics[scale=0.29]{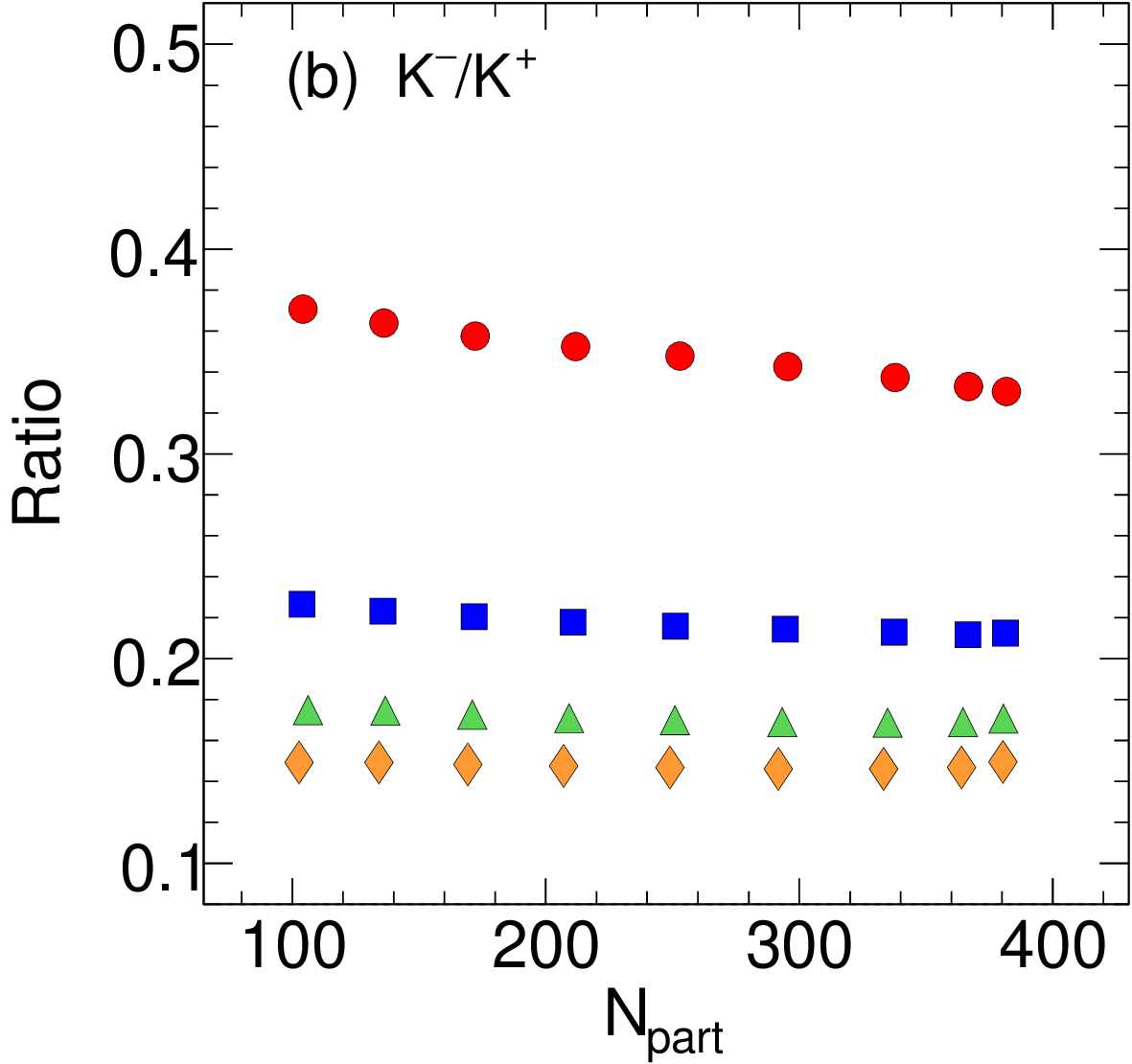}
\includegraphics[scale=0.29]{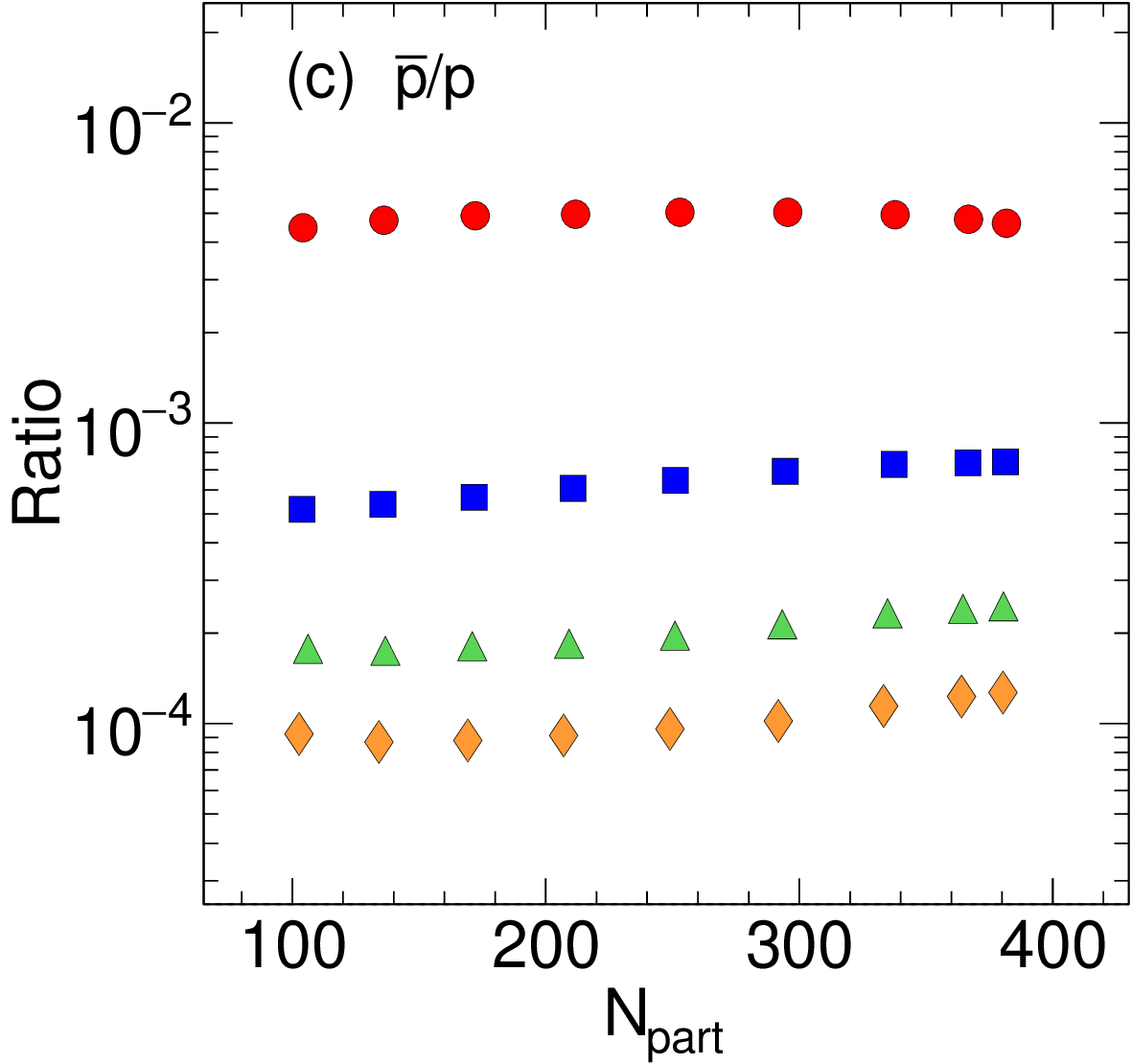}
\caption{(Color online) Antiparticle to particle ratios, (a) $\pi^{-}/\pi^{+}$, (b) $K^{-}/K^{+}$, and (c) $\bar{p}/p$ as a function of $\npart$ at $\midr$ in Au + Au collisions at $\Elabs$ from the PHSD model.}
\label{fig:ratios}
\end{center} 
\end{figure*}

In this sub-section, we discuss various ratios of antiparticle to particle yields and mixed particle yields at $\midr$. Figure~\ref{fig:ratios} shows the ratios, $\ratios$, as a function of $\npart$ in Au + Au collisions at $\Elabs$ from the PHSD model. The $\pi^{-}/\pi^{+}$ ratio do not show significant dependence on collision centrality within the uncertainties over the energy range considered. However, the ratio seems to be larger at the lowest beam energy of 6.7 A~GeV compared to the other energies. This could be attributed to isospin effects and to increased contributions from resonance particles decaying into $\pi^{-}$ and $\pi^{+}$. This observation is consistent with the findings from the STAR experiment at RHIC, which reported similar trends in the beam energy range of $\sqrt{s_{NN}} =$ 7.7 to 200 GeV~\cite{r8}.

The ratio, $K^{-}/K^{+}$, exhibits a clear increase with increasing beam energy, which may be attributed to the growing contribution from pair production processes. At lower beam energies, kaon production is influenced by the interplay between pair production and associated production in strangeness, with the latter being the dominant mechanism. The rapid decrease in the $K^{-}/K^{+}$ ratio below beam energy of 25 A GeV suggests a transition from meson-driven to baryon-driven production mechanism~\cite{r41b}. Furthermore, the observed enhancement of the $K^{-}$ yield in dense baryonic matter could also be due to in-medium modifications of kaon properties, such as reduced effective masses and hadronic interactions in the PHSD model. The $\bar{p}/p$ ratio exhibits no significant centrality dependence in the beam energy range of $E_{lab} =$ 6.7 to 25 A GeV. However, the ratio increases with beam energy, reflecting a higher net-baryon density resulting from greater baryon stopping at lower collision energies within the PHSD model framework.

\begin{figure*}[!htbp]
\begin{center}
\includegraphics[scale=0.3]{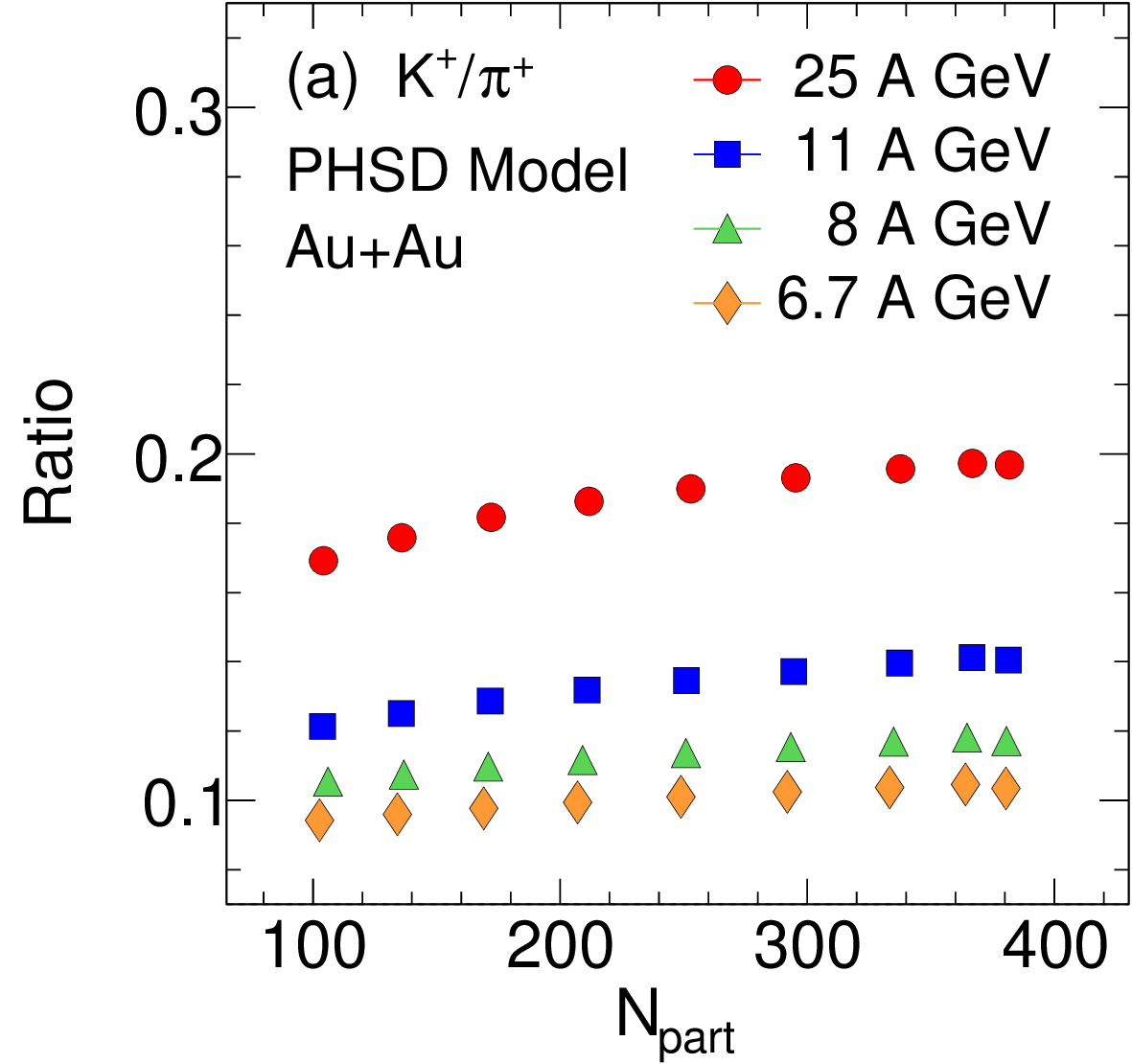}
\includegraphics[scale=0.3]{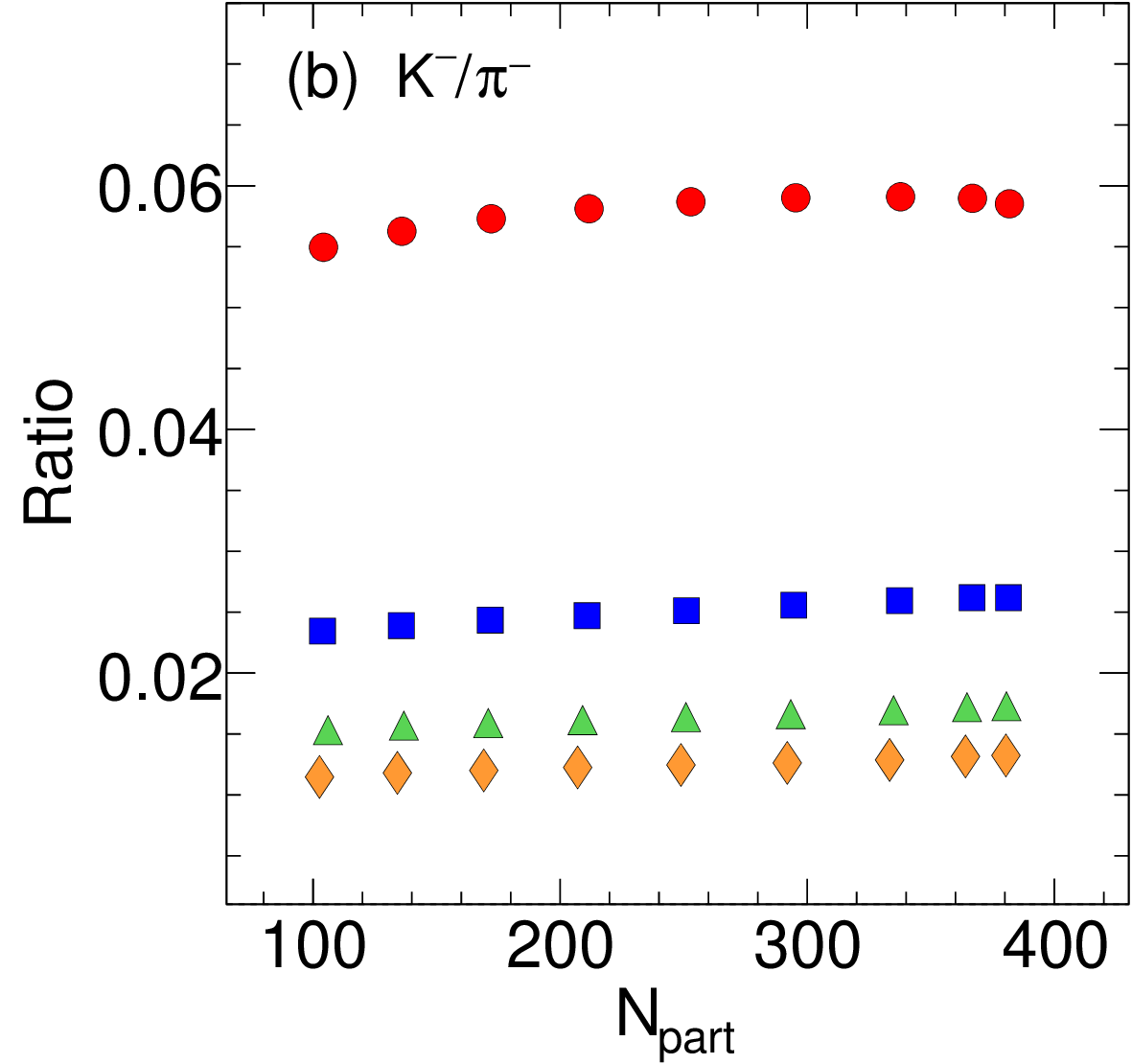} \\
\includegraphics[scale=0.3]{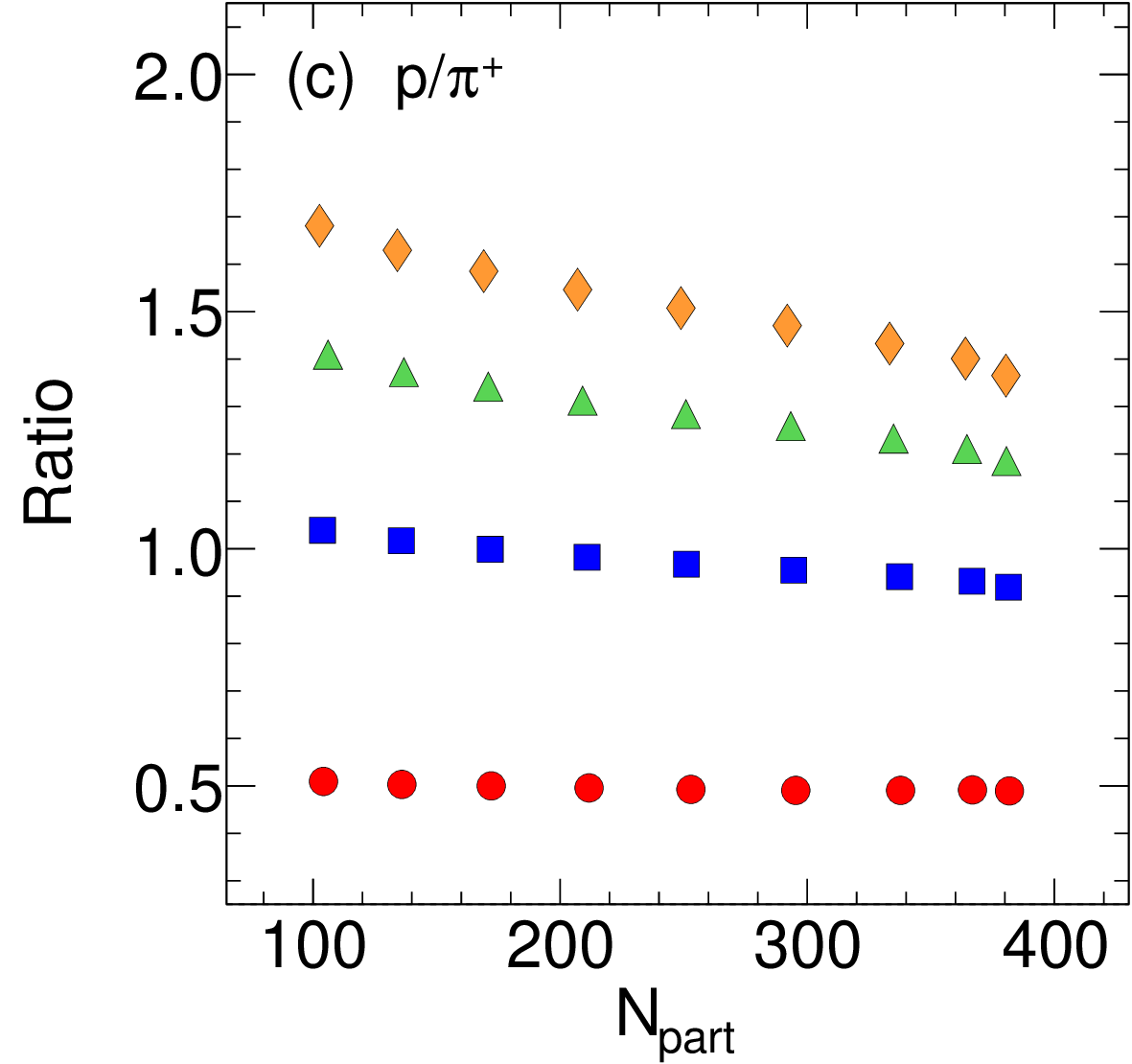}
\includegraphics[scale=0.3]{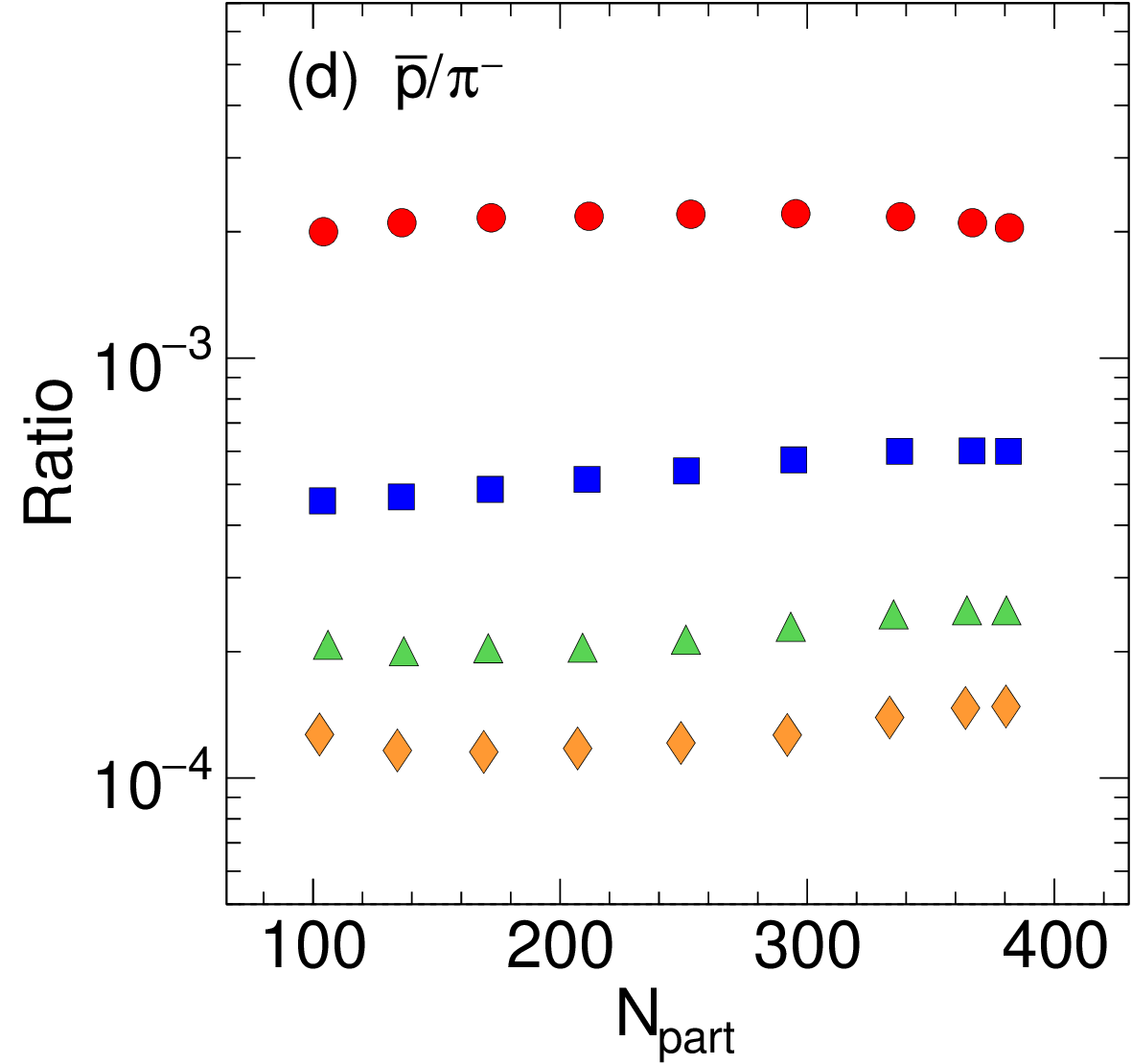}
\caption{(Color online) Particle ratios, (a) $K^{+}/\pi^{+}$ and (b) $K^{-}/\pi^{-}$, (c) $p/\pi^{+}$, and (d) $\bar{p}/\pi^{-}$ as a function of $\npart$ at $\midr$ in Au + Au collisions at $\Elabs$ from the PHSD model.}
\label{fig:ratioKPrPi}
\end{center} 
\end{figure*}

The mixed particle ratios, $K^{+}/\pi^{+}$, $K^{-}/\pi^{-}$, $p/\pi^{+}$, and $\bar{p}/\pi^{-}$ at $\midr$ as a function of $\npart$ in Au + Au collisions at beam energies $\Elabs$ from the PHSD model are presented in Fig.~\ref{fig:ratioKPrPi}. The ratios $K^{+}/\pi^{+}$, $K^{-}/\pi^{-}$, and $\bar{p}/\pi^{-}$ exhibit a weak collision centrality dependence for all beam energies, where as $p/\pi^{+}$ show a significant centrality dependence at lower beam energies. Additionally, the $K/\pi$ ratio shows a significant increase with increasing beam energy, indicating an enhanced contribution to strangeness production. The ratio $\bar{p}/\pi^{-}$ also increases with increasing beam energy, however the ratio $p/\pi^{+}$ decreases with increasing beam energy, showing an opposite trend to the $\bar{p}/\pi^{-}$. This opposite trend could be attributed to high baryon stopping at lower beam energies, which leads to relatively higher proton production compared to pions.

\subsection{Beam energy dependence of particle yields}
A systematic study of the beam energy dependence of particle yields in heavy-ion collisions provides essential insight into baryon stopping, strangeness production, and the thermodynamic properties of strongly interacting matter at finite baryon chemical potential. Low beam energies are associated with high values of $\mu_{B}$, where our theoretical understanding is limited and experimental uncertainties are substantial. This makes systematic measurements of particle yields essential for constraining the equation of state of dense nuclear matter and for exploring possible phase transition. At low and intermediate beam energies, the system evolves through a high net-baryon density regime, where baryon transport strongly influences the chemical composition of the medium, as explored by the heavy-ion collision experiments at the AGS, SPS, and RHIC~\cite{r8,r9,r10,r11,r12,r41b,r42,r43,r44,r45,r46,r47}.

\begin{figure*}
\begin{center}
\includegraphics[scale=0.29]{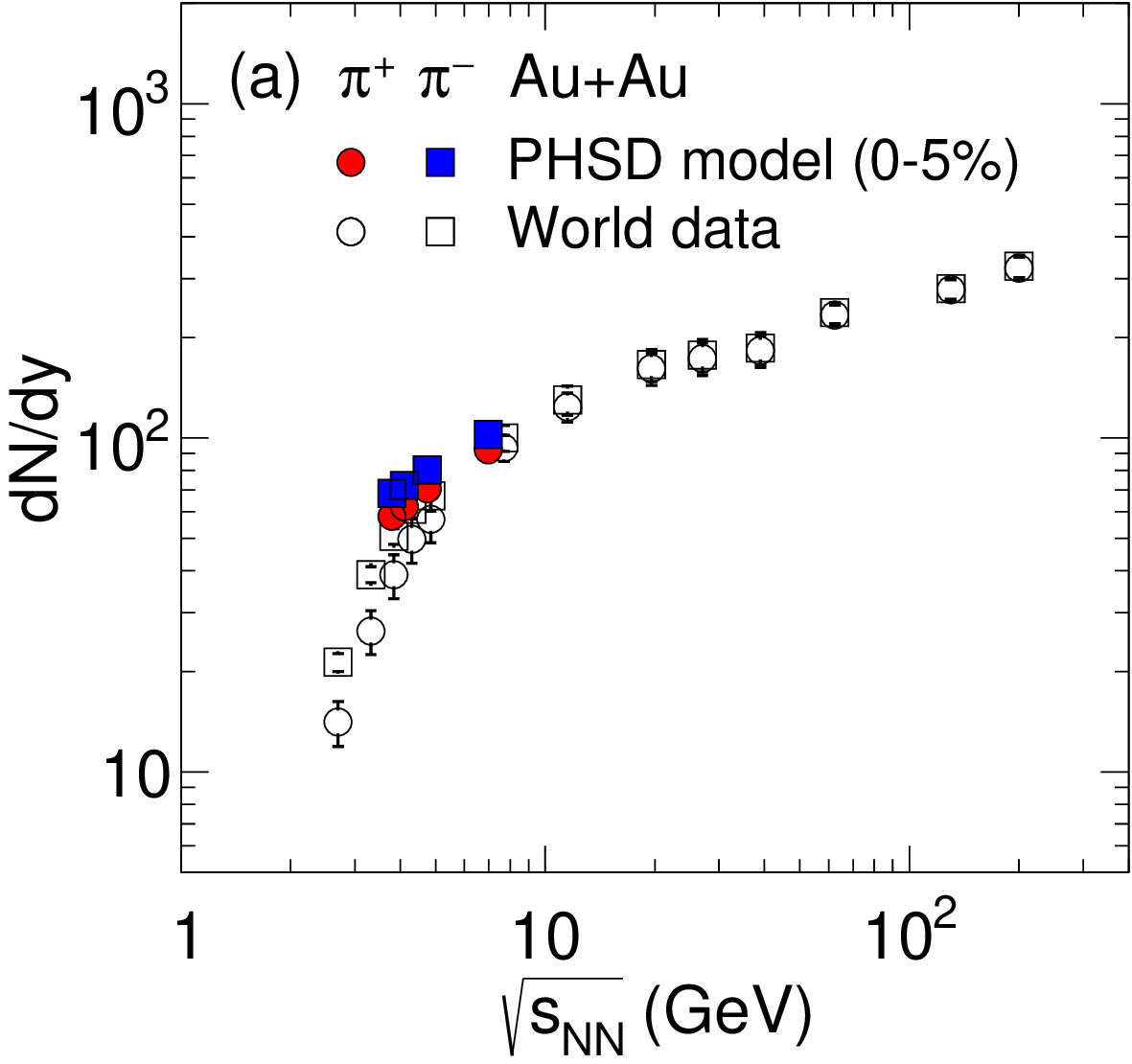}
\includegraphics[scale=0.29]{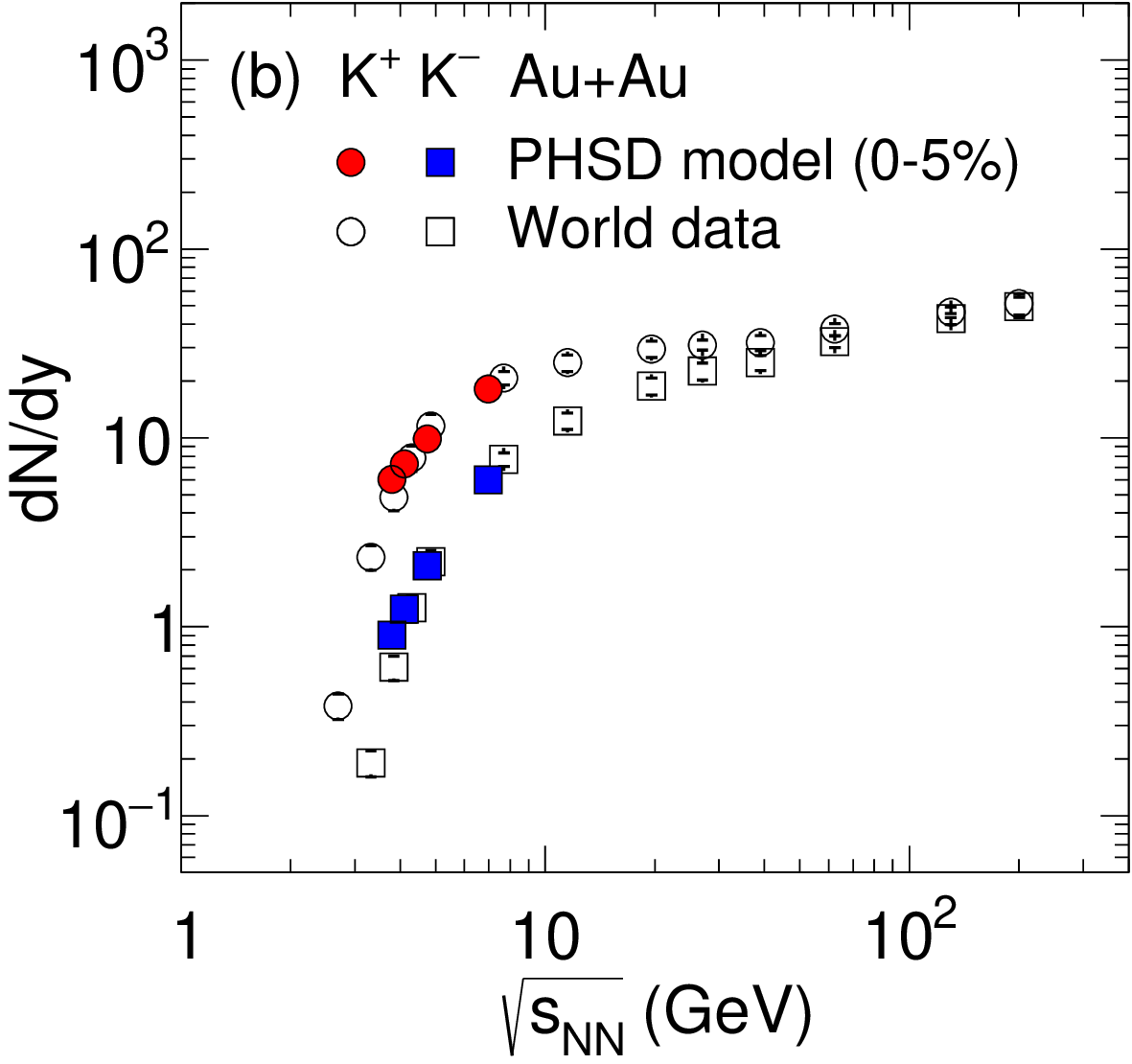}
\includegraphics[scale=0.29]{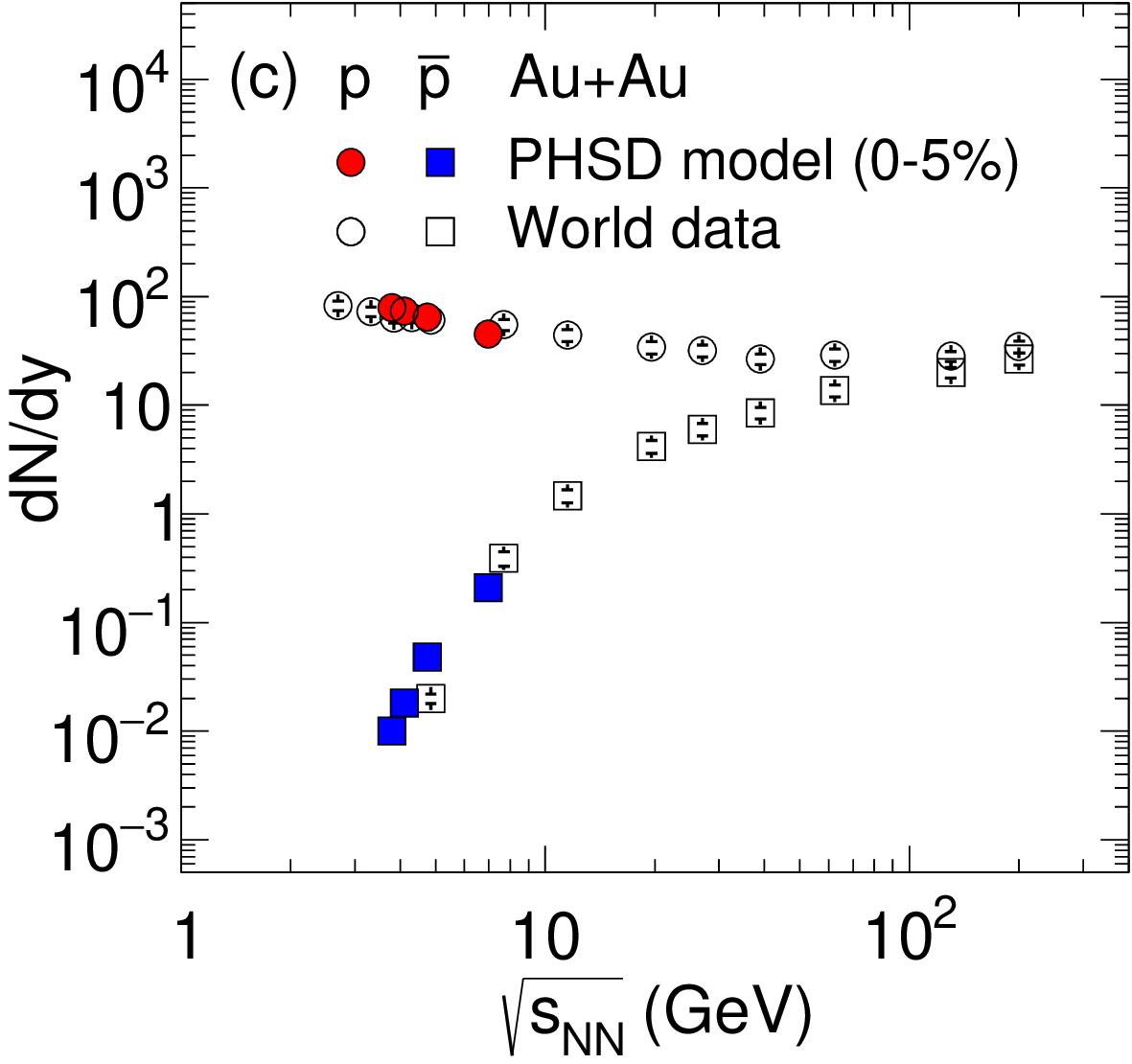}
\caption{(Color online) The mid-rapidity ($|y|<0.5$) yields, $dN/dy$ of (a) $\pi^{\pm}$, (b) $K^{\pm}$, and (c) $p$ and $\bar{p}$ as a function of $\snn$ in 0–5\% central Au + Au collisions from the PHSD model. The results of experimental measurements in Au + Au collisions from AGS~\cite{r41b,r42,r43,r44,r45,r46,r47} and RHIC~\cite{r8,r9,r10,r11,r12} are shown as world data for the comparison.}
\label{fig:yieldsNN}
\end{center} 
\end{figure*}

Figure~\ref{fig:yieldsNN} presents the integrated yields, $dN/dy$, of identified hadrons ($\Ihadrons$) at $\midr$ as a function of $\snn$ in 0–5\% central Au + Au collisions from the PHSD model. The results are compared with measurements from experiments at the RHIC~\cite{r8,r9,r10,r11,r12} and AGS~\cite{r41b,r42,r43,r44,r45,r46,r47}. The integrated yields of $\Ihadrons$ from the PHSD model show a similar dependence on beam energy and are in good agreement with the experimental data. The $dN/dy$ values of $\pi^{\pm}$, $K^{\pm}$, and $\bar{p}$ decrease with decreasing beam energy, whereas the proton yield increases with decreasing beam energy, reflecting the growing contribution of initial-state baryons at mid-rapidity at these low beam energies. 

The $dN/dy$ of $\pi^{+}$ and $\pi^{-}$ increases with increase in beam energy as shown in Fig.~\ref{fig:yieldsNN}(a), and becomes nearly equal at higher energies. The energy dependence of pion yields predicted by the PHSD model suggests an increasing contribution of resonance production, therefore identifying this energy region as a possible transition domain for nuclear matter~\cite{r43}. Figure~\ref{fig:yieldsNN}(b) shows $dN/dy$ of $K^{+}$ and $K^{-}$ as a function of $\snn$ from the PHSD model. The $K^{+}$ yield is more than the $K^{-}$ yield at the studied beam energies. This difference can be attributed to the dominance of associate processes in strangeness production. In contrast, at higher energies, pair production becomes significant~\cite{r39}.  

The energy dependence of antiproton production at these energies provides crucial constraints on the dynamics of the baryon-rich medium. The mid-rapidity $\bar{p}$ production in central Au + Au collisions from the PHSD model shows strong suppression below 25~A GeV in Fig.~\ref{fig:yieldsNN}(c). The energy dependence is in good agreement with the experimental data, suggesting in-medium absorption of antiprotons. Additionally, this highlights the sensitivity of antibaryon yields to the dense baryonic medium~\cite{r41b}.

\subsection{Beam energy dependence of $<\pt>$}
\begin{figure*}
\begin{center}
\includegraphics[scale=0.29]{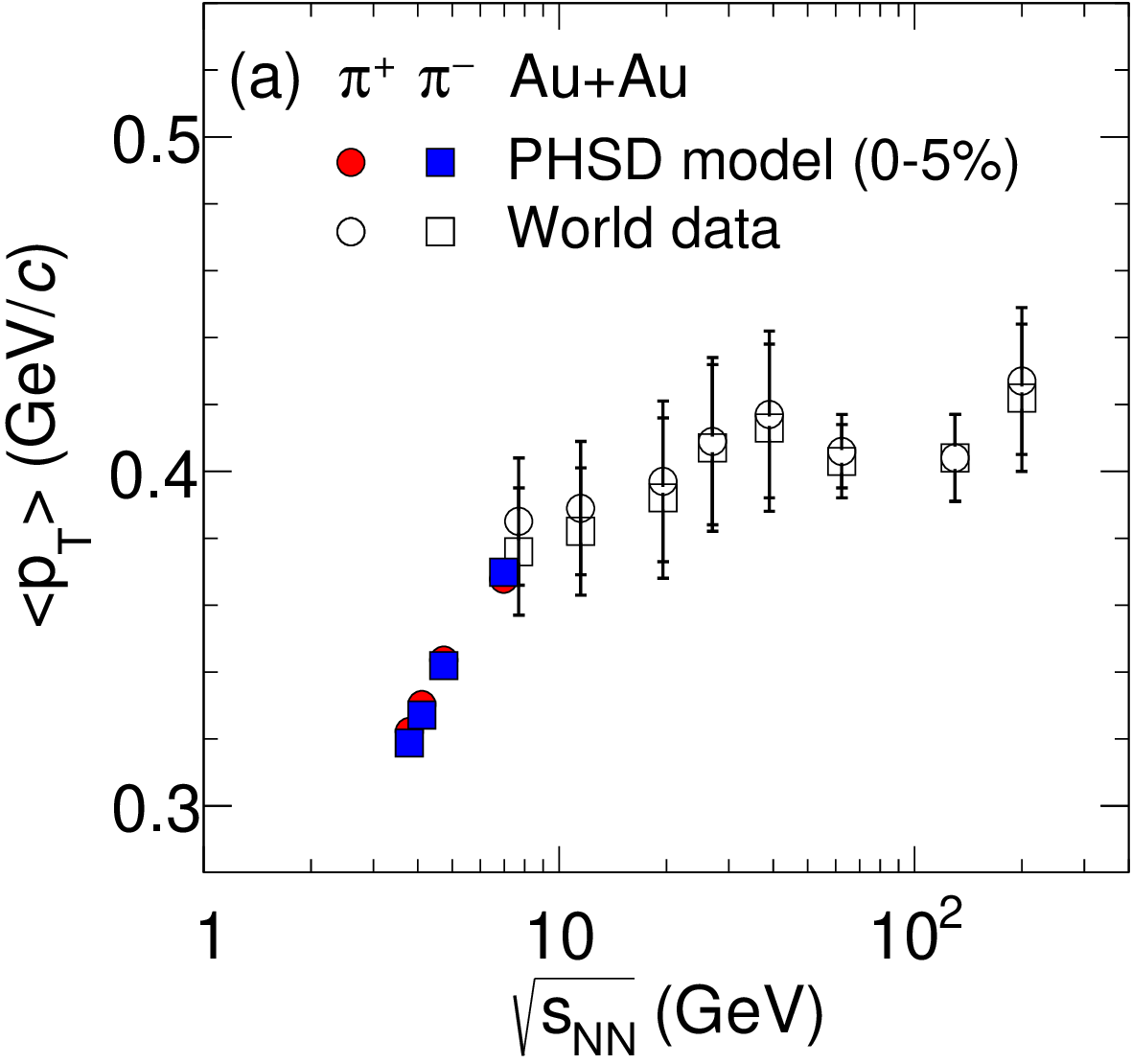}
\includegraphics[scale=0.29]{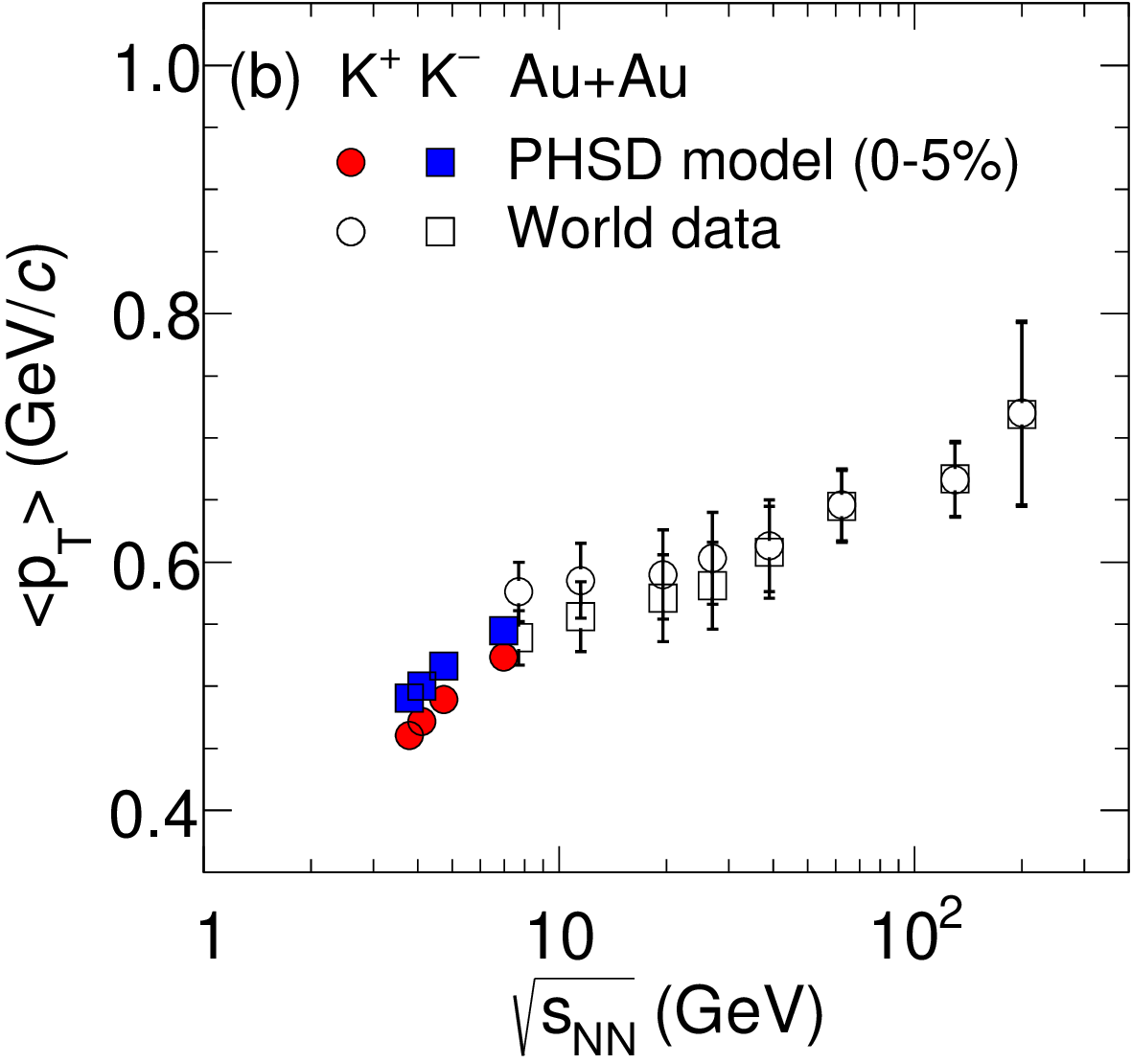}
\includegraphics[scale=0.29]{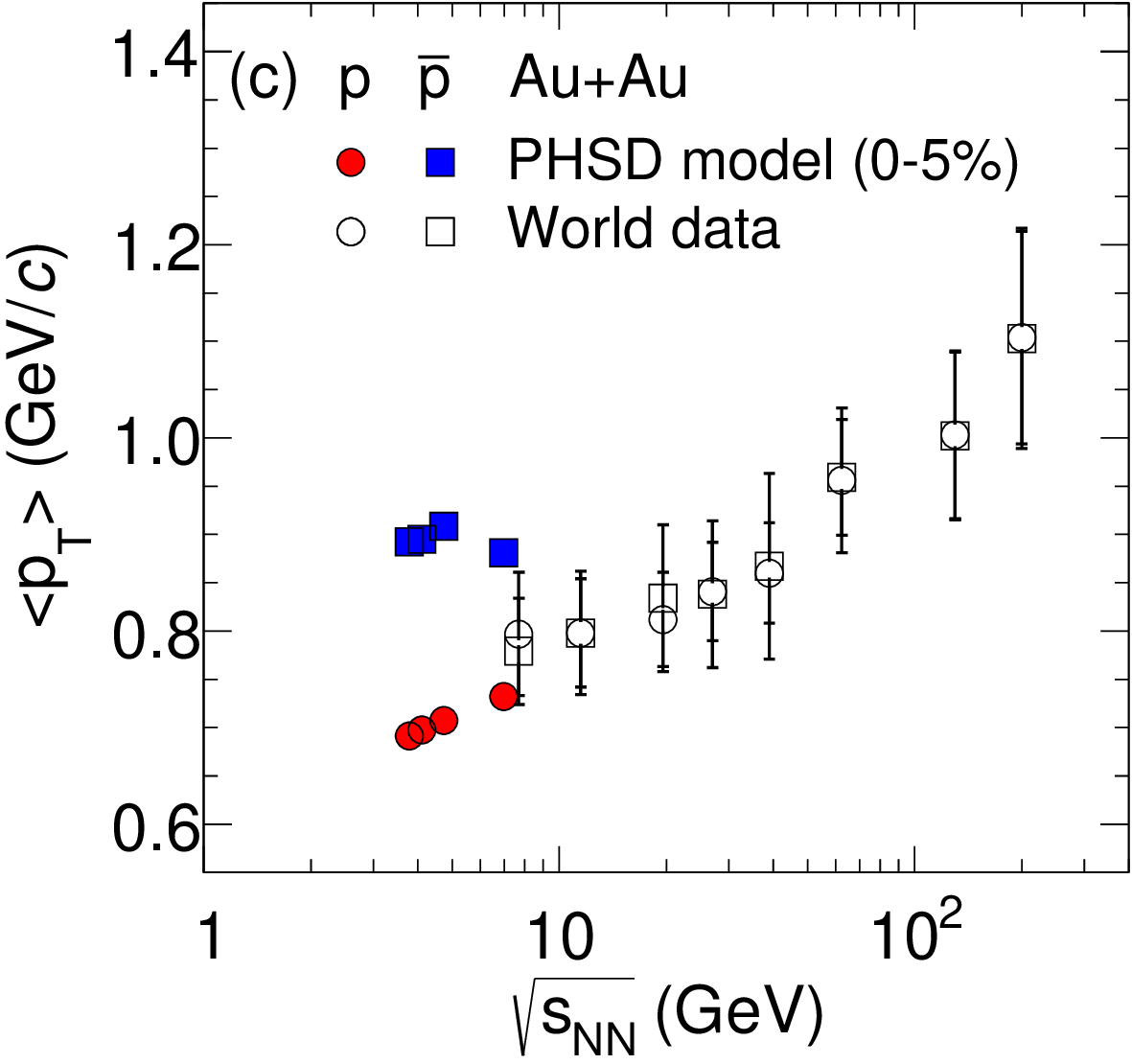}
\caption{(Color online) $\meanpt$ as a function of $\snn$ for (a) $\pi^{\pm}$, (b) $K^{\pm}$, and (c) $p$ and $\bar{p}$ in 0–5\% central Au + Au collisions from the PHSD model. The results of experimental measurements in Au + Au collisions from RHIC~\cite{r8,r9,r10,r11,r12} are shown as world data for the comparison.} 
\label{fig:meanptsNN}
\end{center} 
\end{figure*}
Figure~\ref{fig:meanptsNN} shows the energy dependence of $\meanpt$ for $\Ihadrons$ in 0-5\% central Au + Au collisions from the PHSD model. These results are compared with experimental measurements from RHIC in Au + Au collisions~\cite{r8,r9,r10,r11,r12}. The PHSD results indicate that $\meanpt$ increases with $\sqrt{s_{NN}}$ for pions, kaons, and protons, reflecting the gradual enhancement of transverse collective dynamics as collision energy increases. The beam-energy dependence of $\langle p_T \rangle$ for protons and antiprotons, as obtained from the PHSD model, exhibits a clear splitting. Specifically,  the $\langle p_T \rangle$ for ${\bar{p}}$ is systematically higher than that of $p$ across all studied energies. This observed behavior can be understood in terms of baryon–antibaryon ($B\bar{B}$) annihilation process in a baryon-rich medium. At low and intermediate beam energies, significant baryon stopping results in a high net-baryon density at mid-rapidity, which enhances the probability of annihilation of low-$p_T$ antiprotons with the abundant baryons. As a result, the low-$p_T$ component of the $\bar{p}$ spectra is significantly suppressed, while the high-$p_T$ part remains largely unaffected. This selective depletion results in a hardening of the antiproton $p_T$-spectra, leading to higher $\langle p_T \rangle$ values compared to protons. In contrast, protons are influenced by substantial contributions from transported (initial-state) baryons, which populate the low-$p_T$ region, resulting in comparatively softer $p_T$-spectra. The difference in spectra decreases with increasing $\sqrt{s_{NN}}$, consistent with the reduction in net-baryon density and weaker annihilation effects at higher energies. These findings highlight the role of baryon–antibaryon annihilation as an important mechanism governing particle production in the high baryon density regime.

\section{Summary and discussions}
\label{sec:conclusion}
In this work, we present a study of identified hadron production at $\midr$ in Au + Au collisions over a wide range of beam energies using the PHSD model. These energies overlap with the proposed beam energy range of the CBM experiment ($E_{lab} =$ 6.7, 8, and 11 A GeV). The yields of pions, kaons, and anti-protons decrease with decreasing beam energy at a given centrality, while proton yields increase, highlighting the increasing contribution of baryon stopping at mid-rapidity at lower energies. The production of $K^{-}$ and $\bar{p}$ compared to other charged particles is strongly suppressed at beam energies below 25~A GeV, suggesting the difference between particles composed of only produced quarks and those containing both produced and transported quarks from colliding nucleons. The $\meanpt$ of pions, kaons, and protons decreases with beam energies, reflecting a reduction in collective dynamics at lower energies.

The $\pi^{-}/\pi^{+}$ ratio shows an enhancement at the lowest beam energies attributable to the isospin effects and resonance decays. The $K^{-}/K^{+}$ ratios increase with energy, reflecting the reduction of associated production and dominance of pair production. The ratio $\bar{p}/p$ increases with beam energy, showing a decreased contribution from baryon stopping within the PHSD model framework. The $K/\pi$ ratios increase with beam energy, reflecting the enhanced strangeness production at higher energies. In contrast, the $p/\pi^{+}$ ratio decreases with increasing energy, showing pair production processes dominate over baryon stopping. The $\bar{p}/\pi^{-}$ ratio shows an increase with increasing beam energy. The integrated yields of particles as a function of $\snn$ from the PHSD model are found to be in good agreement with the available measurements from the AGS and RHIC. The $\meanpt$ increases with $\sqrt{s_{NN}}$ for pions, kaons, and proton and exhibits a clear mass ordering at all energies, reflecting the gradual strengthening of transverse collective dynamics with increasing beam energy. The beam-energy dependence of $\langle p_T \rangle$ for protons and antiprotons exhibits a clear splitting, with $\langle p_T \rangle$ of ${\bar{p}}$ systematically higher than $\langle p_T \rangle$ of $p$ at all studied energies. This behavior is attributed to the baryon–antibaryon ($B\bar{B}$) annihilation in a baryon-rich medium.

In summary, the PHSD model provides a qualitatively consistent and comprehensive description of hadron production in Au + Au collisions across a broad range of collision energies ($\Elabs$) and centralities (0-5\% to 70-80\%). The observed systematic trends in $\pt$ spectra, particle yields, mean transverse momentum, and particle ratios highlight the crucial interplay among baryon stopping, strangeness production, baryon-antibaryon annihilation, and associated versus pair-production processes. These results offer valuable theoretical predictions for ongoing and future experimental studies at FAIR and NICA and contribute to a deeper understanding of the properties of strongly interacting matter at high baryon density.

\section{Acknowledgements}
\label{acknowledgement}
T.B. acknowledges the financial support from the Council of Scientific \& Industrial Research (CSIR), Government of India, under the Research Associateship (RA) program, Grant No. 09/0135(23801)/2025-EMR-I. L.K. acknowledges the financial support from Research Grant No. SR/MF/PS-02/2021-PU (E-37120) of the Department of Science and Technology, Government of India. S.K. acknowledges the partial financial support of ANID FONDECYT regular No. 1230987 and ANID CCTVal CIA2500027. This research was supported in part by the cluster computing resource provided by the IT Division at the GSI Helmholtzzentrum für Schwerionenforschung, Darmstadt, Germany.


\begin{thebibliography}{100}
\small
\bibitem{r1} J. Cleymans and K. Redlich, Phys. Rev. C {\bf 60}, 054908 (1999).
\bibitem{r2} F. Becattini, J. Manninen, and M. Gazdzicki, Phys. Rev. C {\bf 73}, 044905 (2006).
\bibitem{r3} A. Andronic, P. Braun-Munzinger, and J. Stachel, Nucl. Phys. A {\bf 772}, 167 (2006).
\bibitem{r4} E. V. Shuryak, Phys. Lett. B {\bf 78}, 150 (1978).
\bibitem{r5} J. Cleymans, R. V. Gavai, and E. Suhonen, Phys. Rept. {\bf 130}, 217 (1986).
\bibitem{r6} F. Karsch, Nucl. Phys. A {\bf 698}, 199c (2002).
\bibitem{r7} S. A. Bass {\it et al.}, Nucl. Phys. A {\bf 661}, 205 (1999).
\bibitem{r8} L. Adamczyk {\it et al.} (STAR Collaboration), Phys. Rev. C {\bf 96}, 044904 (2017).
\bibitem{r9} J. Adam {\it et al.} (STAR Collaboration), Phys. Rev. C {\bf 102}, 034909 (2020).
\bibitem{r9b} A. Nandi, L. Kumar, and N. Sharma, Phys. Rev. C {\bf 102}, 024902 (2020).
\bibitem{r10} M. S. Abdallah {\it et al.} (STAR Collaboration), Phys.Lett. B {\bf 831}, 137152 (2022). 
\bibitem{r11} M. I. Abdulhamid {\it et al.} (STAR Collaboration), JHEP {\bf 10}, 139 (2024). 
\bibitem{r12} M. I. Abdulhamid {\it et al.} (STAR Collaboration), Phys. Rev. C {\bf 110}, 054911 (2024). 
\bibitem{r13} J. Adamczewski-Musch {\it et al.} (HADES Collaboration), Eur. Phys. J. A {\bf 56}, 259 (2020).
\bibitem{r13b} S. Huth {\it et al.}, Nature {\bf 606}, 276-280 (2022).
\bibitem{r14} T. Ablyazimov {\it et al.} (CBM Collaboration), Eur. Phys. J. A {\bf 53}, 60 (2017).
\bibitem{r15} V. Abgaryan {\it et al.} (MPD Collaboration), Eur. Phys. J. A {\bf 58}, 140 (2022).
\bibitem{r16} W. Cassing {\it et al.}, Phys. Rev. C {\bf 93}, 014902 (2016).
\bibitem{r17} A. Palmese {\it et al.}, Phys. Rev. C {\bf 94}, 044912 (2016).
\bibitem{r18} O. Linnyk {\it et al.}, Prog. Part. Nucl. Phys. {\bf 87}, 50-115 (2016).
\bibitem{r19} T. Song {\it et al.}, Phys. Rev. C {\bf 97}, 064907 (2018).
\bibitem{r20} W. Cassing, Nucl. Phys. A {\bf 791}, 365-381 (2007).
\bibitem{r21} W. Cassing, Nucl. Phys. A {\bf 795}, 70-97 (2007).
\bibitem{r22} S. Juchem, W. Cassing, and C. Greiner, Phys. Rev. D {\bf 69}, 025006 (2004).
\bibitem{r23} S. Juchem, W. Cassing, and C. Greiner, Nucl. Phys. A {\bf 743}, 92 (2004).
\bibitem{r24} S. Borsanyi {\it et al.}, Phys. Lett. B {\bf 730}, 99 (2014).
\bibitem{r25} S. Borsanyi {\it et al.}, Phys. Rev. D {\bf 92}, 014505 (2015).
\bibitem{r26} A. Lang {\it et al.}, Z. Phys. A {\bf 340}, 287 (1991).
\bibitem{r27} B. Nilsson-Almqvist and E. Stenlund, Comput. Phys. Commun. {\bf 43}, 387 (1987).
\bibitem{r28} B. Andersson, G. Gustafson, and H. Pi, Z. Phys. C {\bf 57}, 485 (1993).
\bibitem{r29} T. Sjostrand, S. Mrenna, and P. Z. Skands, JHEP {\bf 05}, 026 (2006).
\bibitem{r30} W. Cassing and E. L. Bratkovskaya, Phys. Rep. {\bf 308}, 65 (1999).
\bibitem{r31} J. Schwinger, Phys. Rev. {\bf 82}, 664 (1951).
\bibitem{r32} O. Linnyk {\it et al.}, Phys. Rev. C {\bf 84},  054917 (2011).
\bibitem{r33} V. D. Toneev {\it et al.}, Phys. Rev. C {\bf 85}, 034910 (2012).
\bibitem{r34} V. P. Konchakovski {\it et al.}, Phys. Rev. C {\bf 85}, 011902 (2012).
\bibitem{r35} O. Linnyk {\it et al.}, Phys. Rev. C {\bf 85}, 024910 (2012).
\bibitem{r36} O. Linnyk {\it et al.}, Phys. Rev. C {\bf 87}, 014905 (2013).
\bibitem{r37} L. Ahle {\it et al.}, Phys. Lett. B {\bf 476}, 1 (2000).
\bibitem{r38} J. Klay {\it et al.}, Phys. Rev. C {\bf 68}, 054905 (2003).
\bibitem{r39} L. Ahle {\it et al.}, Phys. Lett. B {\bf 490}, 53 (2000).
\bibitem{r40} J. Klay {\it et al.}, Phys. Rev. Lett. {\bf 88}, 102301 (2002).
\bibitem{r41} C. Blume and C. Markert, Prog. Part. Nucl. Phys. {\bf 66}, 834 (2011).
\bibitem{r41b} J. Barrette {\it et al.} (E877 Collaboration), Phys. Rev. C {\bf 62}, 024901 (2000).
\bibitem{r42} L. Ahle {\it et al.} (E802 Collaboration), Phys. Rev. C {\bf 58}, 3523 (1998).
\bibitem{r43} L. Ahle {\it et al.} (E802 Collaboration), Phys. Rev. C {\bf 60}, 064901 (1999).
\bibitem{r44} L. Ahle {\it et al.} (E866 and E917 Collaboration), Phys. Lett. B {\bf 476}, 1-8 (2000).
\bibitem{r45} B. B. Back {\it et al.} (E917 Collaboration), Phys. Rev. Lett. {\bf 86}, 1970 (2001).
\bibitem{r46} B. B. Back {\it et al.} (E917 Collaboration), Phys. Rev. Lett. {\bf 87}, 242301 (2001).
\bibitem{r47} J. Klay {\it et al.} (E895 Collaboration), Phys. Rev. C {\bf 68} 054905 (2003).
\end{thebibliography}
\end{document}